\documentclass[10pt,twocolumn,english,british,tightenlines,eqsecnum,
floats,aps,amsmath,amssymb,nofootinbib,superscriptaddress,prd,showpacs,showkeys]
              {revtex4}
\usepackage[latin9]{inputenc}
\usepackage{color}
\usepackage{babel}
\usepackage{amsmath}
\usepackage{amssymb}
\usepackage{graphicx}
\usepackage{esint}
\usepackage[unicode=true,pdfusetitle,
 bookmarks=true,bookmarksnumbered=false,bookmarksopen=false,
 breaklinks=false,pdfborder={0 0 1},backref=false,colorlinks=false]
 {hyperref}

\makeatletter
\@ifundefined{textcolor}{}
{%
 \definecolor{BLACK}{gray}{0}
 \definecolor{WHITE}{gray}{1}
 \definecolor{RED}{rgb}{1,0,0}
 \definecolor{GREEN}{rgb}{0,1,0}
 \definecolor{BLUE}{rgb}{0,0,1}
 \definecolor{CYAN}{cmyk}{1,0,0,0}
 \definecolor{MAGENTA}{cmyk}{0,1,0,0}
 \definecolor{YELLOW}{cmyk}{0,0,1,0}
}

\@ifundefined{date}{}{\date{October 16, 2014}}
\usepackage{babel}

\usepackage{euscript}\usepackage{epsfig}

\usepackage{amsfonts}

\usepackage{fancyhdr}

\makeatother

\begin{document}

\title{Noncommutative minisuperspace, gravity-driven acceleration and kinetic inflation}

\author{S. M. M. Rasouli}

\email{mrasouli@ubi.pt}

\affiliation{Departamento de F\'{i}sica, Universidade da Beira Interior, Rua Marqu\^{e}s d'Avila
e Bolama, 6200 Covilh\~{a}, Portugal}

\affiliation{Centro de Matem\'{a}tica e Aplica\c{c}\~{o}es (CMA - UBI),
Universidade da Beira Interior, Rua Marqu\^{e}s d'Avila
e Bolama, 6200 Covilh\~{a}, Portugal}

\author{Paulo Vargas Moniz}

\email{pmoniz@ubi.pt}

\affiliation{Departamento de F\'{i}sica, Universidade da Beira Interior, Rua Marqu\^{e}s d'Avila
e Bolama, 6200 Covilh\~{a}, Portugal}

\affiliation{Centro de Matem\'{a}tica e Aplica\c{c}\~{o}es (CMA - UBI),
Universidade da Beira Interior, Rua Marqu\^{e}s d'Avila
e Bolama, 6200 Covilh\~{a}, Portugal}

\begin{abstract}
In this paper, we introduce a noncommutative version of the Brans-Dicke (BD)
theory and obtain the Hamiltonian equations of motion
for a spatially flat~Friedmann--Lema\^{\i}tre--Robertson--Walker
universe filled with a perfect fluid.
We focus on the case where the scalar potential
as well as the ordinary matter sector are absent. Then, we investigate gravity-driven
acceleration and kinetic inflation in this noncommutative BD cosmology.
In contrast to the commutative case, in which the scale factor and BD scalar
field are in a power-law form, in the noncommutative case
the power-law scalar factor is multiplied by a dynamical
exponential warp factor. This warp factor depends on the noncommutative parameter
as well as the momentum conjugate associated to the BD scalar field.
%This dynamical
%warp factor appears in the differential equation associated to the BD scale
%factor, and hence, except a few particular cases, solving it analytically is a laborious process.
We show that the BD scalar
field and the scale factor
%and its time derivatives, further the values of the BD
%coupling parameter and the integration
%constant associated to the momentum conjugate of the BD scalar field,
effectively depend on the noncommutative parameter.
For very small values of this parameter, we obtain
an appropriate inflationary solution, which can
overcome problems within BD standard cosmology in a more efficient manner.
%than the corresponding commutative case, but
%also it can
Furthermore, a graceful exit from an early acceleration epoch towards a
decelerating radiation epoch is provided.
%In the standard commutative
%case, this (graceful exit) problem is present in all the accelerations in
%D-branch. In the other words, we show that, differently from the inflationary solutions
%in the commutative case in the scalar tensor theory and the pre-big bang
%cosmology, their corresponding solutions in the noncommutative
%case can obey the requirement for the real inflation.
For late times, due to the presence of the noncommutative parameter, we obtain a zero
acceleration epoch, which can be interpreted as the coarse-grained explanation.
 \end{abstract}

\medskip

\pacs{04.50.Kd, 04.20.Jb, 04.60.Bc, 98.80.Bp}

\keywords{Brans-Dicke theory, noncommutativity,
minisuperspace models, quantum cosmology, inflation}

\maketitle

\section{Introduction}
Among the theories alternative to Einstein's general
relativity, the Brans-Dicke (BD) theory~\cite{BD61} is the simplest and the best known.
In the BD theory, the gravitational constant
%(in GR, $G=M_0^{-2}$where $M_0=10^{19}GeV$ is the constant of Planck mass)
has been assumed to be a
dynamical variable, which is proportional to the inverse of a dynamical scalar
field, namely, the BD scalar field, $\phi$.

In the Jordan frame of the BD theory,
the scalar field couples nonminimally only
with the geometry and does not couple directly with the
matter. Hence, the energy-momentum tensor of
the ordinary matter (all types of matter except the BD
scalar field) obeys the usual conservation law.
Moreover,
 %the mentioned difference of the BD with GR,
there is a free dimensionless adjustable parameter, which is called
the BD coupling parameter and denoted by $\omega$. In spite of the theoretical proposals
in which it has been anticipated that the values of
the BD parameter should be of order unity, observational
measurements have indicated that the lower bound on
$\mid\omega\mid$ is large~\cite{Faraoni.book}.

At a classical level, to obtain results in agreement with observational
data, for an early as well as a late time universe, other extended
versions of the BD theory (scalar-tensor theories) have been applied.
In these theories, in contrary to the standard version of the BD theory,
it has been assumed that either the BD coupling
parameter should be a general function of the BD scalar field~\cite{BP01}, and/or
a scalar potential~\cite{MC07} (which is also a function of $\phi$)
must be added by hand\rlap.\footnote{It has
been recently shown~\cite{RFM14} that, instead of
an {\it ad hoc} assumption, such a scalar potential could be
induced from the geometry of an extra dimension.
Such formalism for an anisotropic Bianchi type
I solution has been examined in~\cite{RFS11-R14}.}
It is also established that the BD theory not only can
provide observational consequences to convince the original aims of the
theory, but also it is possible to construct interesting quantum
cosmological models, which may present appropriate scenarios to
study the inflationary universe~\cite{LS89-SA90-S98-CL11-CFPS12,BM90}.

In the BD setting of~\cite{Lev95,Lev95-2}, which is of
interest in this study, an accelerated expanding universe was~not obtained
by adding a scalar potential or a cosmological constant.
However, contrary to the standard BD theory, a
variable BD coupling parameter rather than a constant one has been assumed. Being more concrete, an accelerated
expanding universe emerges from
%has been identified just with the assistance of
the kinetic
energy density of a dynamical Planck mass\footnote{Throughout this
paper, we will use Planck units. Thus, Planck mass, which is a variable in BD
theory, is given by $m_{\rm Pl}=\phi^{1/2}$.} without
introducing any scalar potential or cosmological constant.
More precisely, in this formalism, the pressure
associated to the kinetic energy density is negative.
%(Whilst, a free minimally coupled field does have always positive pressure.)
%A direct consequence of such a negative pressure is producing an accelerating cosmic expansion
%However, the only deviation from the original BD theory is
%assuming a variable BD coupling parameter instead of a constant one.
%Such an accelerating scale factor is a necessary condition for
%explaining the predictions of the observational data for the early
%as well as late time phase of the universe.

Although having an accelerating scale factor is required to
explain the early as well as late time phases of the
Universe,
%but, as the observational data indicates, obtaining a suitable
%accelerating scale factor at early times needs substantial
additional features need to be satisfied.
%Let us be more concrete.
In fact, the early Universe must inflate such that it can overcome the
problems with the standard cosmology.
Moreover, a successful inflationary model
must exit from an accelerating phase
and proceed to a decelerated expansion.

In Ref.~\cite{Lev95-2}, it has been shown that, to meet sufficient inflation,
it is required to have an accelerating scale factor in the Einstein frame.
However, there is no source to get an accelerating scale
factor in that frame in the model investigated in~\cite{Lev95-2}.
Thus, kinetic inflation, even by assuming a variable BD coupling
parameter, cannot lead to today's Universe.
Namely, in the commutative case of the BD theory (in the Jordan frame), there
is an important problem with kinetic
inflation, even with a variable $\omega$:
regardless of the form of $\omega(\phi)$, all the
D branch\footnote{We will introduce the D and X branches in footnote~6.}
solutions are encountered with the graceful exit problem~\cite{Lev95-2}.
The graceful exit problem is also an obstacle in (accelerated)
 inflation within more general solutions in the context of string theory~\cite{BV94}.

%Levin~\cite{} has shown that the some of
%the mentioned problems, specially the
%graceful exit problem, still squire the the gravity driven
%acceleration models based on the BD theory with a
%variable $\omega$.
In this work, we will present a model
which can give an accelerating scale factor for the early
Universe, without encountering the above-mentioned problems.
%, especially the graceful exit problem.
Moreover, we will show that the nominal as well as
sufficient conditions, which are required for an inflationary epoch,
are satisfied in a more convenient manner when the noncommutativity parameter is present.
% than the proposals
%prepared for the kinetic inflation in the context of more generalized
%models of the standard BD theory, the scalar tensor theories.
Our model will not be constructed by adding a scalar
potential or by taking a variable BD coupling parameter.
Instead, we will study the effects of a
noncommutativity in a cosmology constructed
with a flat~Friedmann--Lema\^{\i}tre--Robertson--Walker (FLRW)
model in the context of the BD theory, in the absence of the ordinary matter.

Noncommutative field theory~\cite{CDS98-SW99-DN01} has
been applied to gravitational models which led
to present a few noncommutative proposals for gravity~\cite{GORS03-GORS03-2-ADMW06-EGOR08}.
Such approaches have indicated that their
corresponding noncommutative field equations are very complicated to solve.
However, by applying some arguments, gravitational models based on
noncommutativity with simplified field equations involving noncommutative effects have been obtained.
Basically, by means of applying an effective
noncommutativity on a minisuperspace, the noncommutative deformations
of the minisuperspace can be investigated at the quantum level.
At the classical level, noncommutative deformations have also
been studied; see, e.g.~\cite{BP04-PM05-AAOSS07-GSS07,GSS11,RFK11,RZMM14}.

The major objective of this paper will be to construct the spatially flat
FLRW field equations for a generalized BD theory by means of the Hamiltonian formalism
in a noncommutativite minisuperspace. Then, we proceed to obtain the
solutions for very special cases and investigate the effects of
noncommutativity.
%We will start to work with a spatially flat FLRW universe in
%the context of an extended version of the BD theory.
%More precisely, by applying the Hamiltonian formalism,
%we derive the equations of motion for the BD
%theory (with a phenomenological scalar potential) in the commutative case. Then,
By introducing a noncommutative Poisson bracket
between the BD scalar field and the logarithm of
scale factor, we will construct a noncommutative BD cosmology.
The effects of such a noncommutativity on the BD vacuum solutions are discussed.

Our paper is, therefore, organized as follows.
In Sec.~\ref{NC-BDT}, the general Hamiltonian
equations of motion for an extended version of a BD theory (in Jordan frame) in
the presence of a special kind of a noncommutativity
for a spatially flat FLRW universe are derived.
In Sec.~\ref{Vacuum-NC-BD},
%to simplify very complicated noncommutative field equations,
we restrict ourselves to solve the field equations for a case in which
there is not a scalar potential or an ordinary matter.
In Sec.~\ref{Kinetic inflation}, we will argue
that the obtained solutions in section~\ref{Vacuum-NC-BD} can be a successful
alternative for a kinetic inflationary model. In Sec.~\ref{Conclusions},
we will summarize and analyze the results of the paper.

\section{Noncommutative Cosmological equations in Brans-Dicke Theory}
\indent
\label{NC-BDT}
%In this section, we review the BD theory in the Hamiltonian
%formalism.
Let us start with the spatially flat FLRW
metric as the background geometry, namely
\begin{equation}\label{metric1}
ds^{2}=-N^2(t)dt^2+e^{2\alpha(t)}\left(dx^2+dy^2+dz^2\right),
\end{equation}
where $N(t)$ is a lapse function and $a(t)=e^{\alpha(t)}$ is the scale factor.
We will work with a Lagrangian density of the BD
theory\footnote{In the Lagrangian density associated to
the original BD theory, there is no scalar
potential~\cite{BD61,Far09}.
However, for simplicity, we entitle the Lagrangian density~(\ref{lag1}) as
the BD Lagrangian density. We should note that, in the next sections,
we will work in the context of the standard BD theory.}
 in the Jordan frame~\cite{BD61,Jordan55-FGN99} as
\begin{eqnarray}\label{lag1}
{\cal L}[\gamma,\phi]&=&\sqrt{-\gamma}\left[\phi R-
\frac{\omega}{\phi}\gamma^{\mu\nu}\nabla_\mu\phi\nabla_\nu\phi-V(\phi)\right]\\\nonumber
&+&\sqrt{-\gamma}{\cal L_{\rm matt}},
\end{eqnarray}
where the greek indices run from zero to $3$, and
${\cal L_{\rm matt}}=16\pi\rho(\alpha)$ (where $\rho$ is the energy density) is the
Lagrangian density associated to the ordinary matter. In order to have an attractive
 gravity, we should notice that the BD scalar field $\phi$ must take positive values. $V(\phi)$
is the scalar potential, and $R$ is the Ricci scalar
associated to the metric $\gamma_{\mu\nu}$, whose determinant was denoted by $\gamma$.
In this work, we will
assume the BD coupling parameter $\omega$ to be a constant
and, in vacuum, requiring stability in Lorentzian space, it must be restricted as
$\omega>-3/2$~\cite{C98,BKM04-DDB07-BS07-B09}.
%for non--ghost scalar field
%situations~\cite{BKM04,DDB07,BS07,B09}.
It is straightforward to show that the Hamiltonian of the model is given by
\begin{eqnarray}\label{Ham-1}
{\cal H}\!\!\!&=&\!\!\!-\frac{Ne^{-3\alpha}}{2(2\omega+3)\phi}
\left(\frac{\omega}{6}P_\alpha^2-\phi^2P_\phi^2+\phi P_\alpha P_\phi\right)\\\nonumber
\!\!\!&+&\!\!\!Ne^{3\alpha}\left(V-16\pi\rho\right),
\end{eqnarray}
where $P_\alpha$ and $P_\phi$ are the
conjugate momenta associated to the $\alpha$ and $\phi$, respectively.
We will be working with
the comoving gauge; namely, we have set $N(t)=1$.
Thus, by applying the above Hamiltonian, the equations of motion
corresponding to the phase space coordinates
$\{\alpha,\phi;P_\alpha,P_\phi\}$, in which the Poisson algebra is
$\{\alpha,\phi\}=0$, $\{P_\alpha,P_\phi\}=0$,
$\{\alpha,P_\alpha\}=1$ and $\{\phi,P_\phi\}=1$, are given by
\begin{eqnarray}\label{diff.eq1}
\dot{\alpha}\!\!&=&\!\!-\frac{e^{-3\alpha}}{2(2\omega+3)\phi}\left(\frac{\omega}
{3}P_\alpha+\phi P_\phi\right),\\
  \label{diff.eq2}
\dot{P_\alpha}\!\!&=&\!\!e^{3\alpha}\left[-6V+16\pi\left(6\rho+\frac{d \rho}{d \alpha}\right)\right],\\
\label{diff.eq3}
\dot{\phi}\!\!&=&\!\!-\frac{e^{-3\alpha}}{2(2\omega+3)}\left(P_\alpha-2\phi P_\phi\right),\\
 \label{diff.eq4}
\dot{P_\phi}\!\!&=&\!\!\frac{e^{-3\alpha}}{2(2\omega+3)\phi}\left(P_\alpha-2\phi P_\phi\right)P_\phi\\\nonumber
 \!\!\!&-&\!\!\!\frac{e^{3\alpha}}{\phi}\left(V+\phi\frac{dV}{d\phi}-16\pi\rho\right),
  \end{eqnarray}
where a dot denotes the differentiation with respect to the cosmic time.
Because of the homogeneous and isotropic FLRW
universe choice, we have assumed that the spatial gradients in
the BD scalar field are negligible, namely,
%the BD scalar depends only on the cosmic time.
$\phi=\phi(t)$. By using the equation of state associated
to a perfect fluid and the Hamiltonian constraint,
it is straightforward to derive the usual FLRW field equations in the context of the BD cosmology.
However, in this paper, we prefer to work with the first order Hamiltonian differential equations.
%As the BD equations of motion for the above model
%have been given in the literature, let us before deriving them

We will investigate the effects of noncommutativity in this
cosmological model.
In fact, in order to achieve the
corresponding FLRW equations for a noncommutative
setting, we should begin from a noncommutative
theory of gravity. However, as performing such a
procedure is a complicated process, it is
%a profound and unwieldy obligation,
usually replaced by an
effective noncommutativity in the minisuperspace~\cite{COR02,BP04-PM05-AAOSS07-GSS07}.
%Following the mentioned effective noncommutativity
%in the context of quantum cosmology,
By modifying the Poisson
algebra, some particular noncommutative frameworks have
been applied to a minimally coupled scalar field cosmology~\cite{GSS11}, namely, in quantum cosmology.
In particular,
%inspired with the mentioned
%approaches, based on a physical motivations, a particular
a dynamical deformation between the momenta
associated to the scale factor and scalar field has been used in
both of nonminimally and minimally coupled scalar field
cosmology to discuss the corresponding effects in the
evolution of the Universe and singularity formation~\cite{RFK11,RZMM14}.

In order to investigate the effects of a classical
evolution of the noncommutativity on the
cosmological equations of motion in the BD theory, we propose the
following Poisson commutation relations between
the variables:
\begin{eqnarray}\label{NC-Poisson}
\{\alpha,\phi\}=\theta,\hspace{10mm} \{P_\alpha,P_\phi\}=0,\\\nonumber
\{\alpha,P_\alpha\}=1,\hspace{10mm} \{\phi,P_\phi\}=1,
\end{eqnarray}
where the noncommutative parameter $\theta$ is a constant.
Applying the commutation relations~(\ref{NC-Poisson}) leads us to the following
deformed equations of motion
\begin{eqnarray}\nonumber
\dot{\alpha}\!\!\!&=&\!\!\!-\frac{e^{-3\alpha}}{2(2\omega+3)\phi}
\left[\frac{\omega}{3}P_\alpha+\phi P_\phi
\!+\!\theta(P_\alpha-2\phi P_\phi)P_\phi\right]\\\label{NC.H.eq1}
&+&\theta\left(\frac{e^{3\alpha}}{\phi}\right)
\left[V(\phi)+\phi\frac{dV(\phi)}{d\phi}-16\pi\rho\right],\label{NC.H.eq1}\\\nonumber
 \dot{\phi}\!\!&=&\!\!-\frac{e^{-3\alpha}}{2(2\omega+3)}\left(P_\alpha-2\phi P_\phi\right)\\
&-&6\theta e^{3\alpha}\left[V(\phi)
-16\pi\left(\rho+\frac{1}{6}\frac{d\rho}{d\alpha}\right)\right],\label{NC.H.eq2}
\end{eqnarray}
where, as the equations of motion associated to the momenta $P_{\rm a}$ and $P_{\rm \phi}$
under the proposed noncommutative deformation do not change,
we have abstained from rewriting them. Equations~(\ref{NC.H.eq1}) and (\ref{NC.H.eq2})
together with those for the momenta, namely, Eqs.~(\ref{diff.eq2}) and (\ref{diff.eq4}),
are the Hamiltonian equations for the noncommutative BD
setting, and obviously, the standard commutative equations
are recovered in the limit $\theta\rightarrow0$.

In the next
sections, we investigate the cosmological implications of this
model for a very simple case in which the scalar potential and the ordinary matter are absent.

\section{Gravity-Driven acceleration for cosmological models in the commutative and noncommutative BD theory}
\indent
\label{Vacuum-NC-BD}
Let us assume a very simple case in which we set $\rho=0$ and $V(\phi)=0$.
In the commutative setting of BD
theory, such a model has been considered as
an appropriate approach in which the key ideas of
the duality and branch changing have been studied~\cite{L97}.
%The other possibility is finding appropriate
In addition, as mentioned, a gravity-driven acceleration
epoch is obtained without introducing any scalar
potential, cosmological constant, and/or ordinary matter.
In the commutative case, such solutions, by assuming a variable BD coupling parameter, have been
investigated in detail in Refs.~\cite{Lev95,Lev95-2}.
In what follows, we will study the effects of a constant noncommutative parameter
introduced by relation~(\ref{NC-Poisson}) on the behavior of the
cosmological quantities. We should remind
that, in contrast to the approaches of~\cite{Lev95,Lev95-2}, we have assumed the original
BD theory in which $\omega$ should be a constant.
As we will see, the presence of the noncommutative parameter leads us to
some interesting consequences.
% which can be interpreted as
%quantum gravity effects in the early as well as late epoches.

In the absence of the scalar potential and ordinary matter, from (\ref{diff.eq2})
we get $\dot{P_\alpha}=0$, which gives a constant
of motion and may assist us to solve the rest of the equations of motion.
 Thus, we get $P_\alpha=c_1$;
also, Eqs.~(\ref{diff.eq4}) and (\ref{NC.H.eq1}) give $P_{\phi}=c_2^{\pm}\phi^{-1}$ where
$c_1$ and $c_2^{\pm}\neq0$ are the integration constants.
These constants are not independent; by substituting them into the
Hamiltonian constraint, we get the following relation between them:
\begin{eqnarray}\label{c1-c2}
c^{\pm}_1=\frac{3|c_2|}{\omega}\left[-{\rm sgn}(c_2)\pm\xi\right],
\end{eqnarray}
where  $\omega\neq0$, $\xi\equiv\sqrt{1+2\omega/3}$, and ${\rm sgn}(x)=x/|x|$ is the signum function.
Thus, from (\ref{NC.H.eq2}), $\dot{\phi}$ is
written as\begin{eqnarray}\label{phi-dot}
\dot{\phi}=-\frac{f^{\pm}}{\xi a^3}\hspace{5mm}{\rm where}\hspace{5mm}
f^{\pm}\equiv\frac{|c_2|}{2\omega}\left[-{\rm sgn}(c_2)\xi\pm1\right].
\end{eqnarray}
By employing the obtained expressions
associated to the momenta and the
integration constants, Eqs.~(\ref{NC.H.eq1}) and (\ref{NC.H.eq2}) lead us to
\begin{eqnarray}\label{H}
H=h^{\pm}\left(\frac{\dot{\phi}}{\phi}\right)\hspace{5mm}{\rm where} \hspace{5mm}
 h^{\pm}\equiv g^{\pm}+\frac{c_2\theta}{\phi},
\end{eqnarray}
$H=\dot{a}/a$ is the Hubble constant and $g^{\pm}$ is given by
\begin{eqnarray}\label{g}
g^{\pm}\equiv-\frac{1}{2}\left[1\pm {\rm sgn}(c_2)\xi\right].
\end{eqnarray}
Notice that the above equations corresponding with
each sign of\footnote{For simplicity of expressing the
quantities, we will sometimes drop the index $\pm$.} $c_2$
give two branches for the Hubble parameter.
As we have assumed $\phi>0$, i.e., an attractive
gravity~\cite{Faraoni.book}, in order to discuss an expansion or contraction,
 the values of $\dot{\phi}$ as well as $h$
 corresponding to each branch\footnote{Following~\cite{BV94,Lev95,Lev95-2}, for the commutative
  case, we will call the branches as follows.
  Although in the Jordan frame, there are some solutions in which the scale
  factor decreases, we can still obtain an expanding universe for both of the branches. However, in the Einstein
  frame, one of the branches always leads to an expanding universe, while
  the other gives a contracting universe.
  Therefore, the solutions correspond to the former, and the
  latter are called the X branch and D branch, respectively.
  Throughout our paper, when $c_2>0$, the X branch solutions
  correspond to the upper sign, while the D branch solutions
  correspond to the lower sign.
   For the case where $c_2<0$, we should note
    the transformations obtained after Eq.~(\ref{duality}).}
   must be considered.
  For instance, by considering a special case by supposing $c_2>0$ and $\theta=0$,
  we get $H=-1/2\left(\dot{\phi}/\phi\right)(1\pm\xi)$.
  In this case, for $\xi<1$, $H>0$ only when $\dot{\phi}<0$,
   and $H<0$ only when $\dot{\phi}>0$ (for both of the branches).
  While, for $\xi>1$, to have a positive Hubble
  expansion we must choose the upper sign for $\dot{\phi}<0$ and the lower sign for $\dot{\phi}>0$.

  Let us take a general case. We obtain the acceleration of the scale factor as
   \begin{eqnarray}
\frac{\ddot{a}}{a}&=&H^2+\dot{H}=-\frac{1}{6\phi}\left[\rho^{(\phi)}+3p^{(\phi)}\right]\\\nonumber
&=&-\left(\frac{\dot{\phi}}{\phi}\right)^2
\left(2h^2+h+\frac{c_2\theta}{\phi}\right),\label{acceleration}
 \end{eqnarray}
  where the energy density and pressure associated to the BD scalar field are given by~\cite{RFM14}
  \begin{eqnarray}\label{rho-phi}
 \rho^{(\phi)}\!\!&\equiv&\!\!-T^{0(\phi)}_0=3h^2\left(\frac{\dot{\phi}^2}{\phi}\right),\\
 p^{(\phi)}\!\!&\equiv&\!\!T^{i(\phi)}_i=\left(3h^2+2h+
 \frac{2c_2\theta}{\phi}\right)\left(\frac{\dot{\phi}^2}{\phi}\right),\label{p-phi}
  \end{eqnarray}
 where $i=1,2,3$ with no sum and we have used relations (\ref{phi-dot}) and (\ref{H}).
 Hence, in order to have an accelerating universe, the following constraint must be satisfied
 \begin{eqnarray}\label{h-con}
 2h^2+h+\frac{c_2\theta}{\phi}<0.
 \end{eqnarray}
 More precisely, while the Universe evolves, if the functional form of
 $h$ [which is given by (\ref{H})] changes such that it obeys the constraint~(\ref{h-con}),
 then the Universe will be in an accelerating phase.
 As $\rho^{(\phi)}>0$, relation~(\ref{p-phi}) and constraint~(\ref{h-con})
 indicate that the pressure will be negative.

In the particular case where $\theta=0$, by using relations~(\ref{H})
and~(\ref{g}), the constraint~(\ref{h-con}) reduces to
\begin{eqnarray}\label{h-con2}
 \xi[\xi\pm {\rm sgn}(c_2)]<0.
 \end{eqnarray}
As $\xi>0$, the only acceptable solution will give a
constraint on the BD coupling parameter,
namely, $\omega<0$ ($\xi<1$), which corresponds to the negative values for $c_2$
when the upper sign is chosen, while by taking positive
values of $c_2$, the lower sign must be chosen.
Namely, for the commutative case, under changing the sign
of $c_2$, the upper and lower solutions can be exchanged.
%For this, let us only investigate these solutions for only the positive values of the $c_2$.//

From (\ref{H}), we can easily obtain a relation
between the scale factor and the BD scalar field as
\begin{eqnarray}\label{NC-a-phi}
a(t)=a_i[\phi(t)]^{g}e^{-c_2\theta\phi^{-1}},
\end{eqnarray}
where $a_{\rm i}=e^{\alpha_{\rm i}}$ is another integration
constant, which is associated to $\alpha$ in a
specific time. As is seen from~(\ref{NC-a-phi}),
the noncommutative parameter shows itself in the
power of an exponential warp factor, which, in turn, is a linear
function of the dynamical momentum associated to
the BD scalar field. This time-dependent warp factor
appears in the differential equation associated
to $\phi$ [see Eq.~(\ref{NC-diff-phi})] and makes it very
complicated, such that we have to solve it numerically instead.
Inserting (\ref{NC-a-phi}) into Eq.~(\ref{phi-dot}) gives a differential equation for
the BD scalar field as
\begin{eqnarray}\label{NC-diff-phi}
\dot{\phi}\phi^{3g}e^{-3c_2\theta\phi^{-1}}=-\frac{f}{a_i^3\xi},
\end{eqnarray}
where, according to~(\ref{phi-dot}), $f$ depends
on the $c_2$ and the BD coupling parameter, $\omega$.

For a general noncommutative case, solving (\ref{NC-diff-phi})
analytically is very difficult\rlap.\footnote{If we integrate
both sides of Eq.~(\ref{NC-diff-phi}) (by assuming a certain
integral over the BD scalar field and the cosmic time),
the integral on the lhs (over $\phi$) gives an upper (or lower) incomplete
Gamma function, namely, $\Gamma(-1-3g,3c_2\theta/\phi)$.}
 Thus, in the following subsections, by means of a few numerical endeavors,
we will analyze the solutions for different cases.

\begin{figure}
\centering\includegraphics[width=3in]{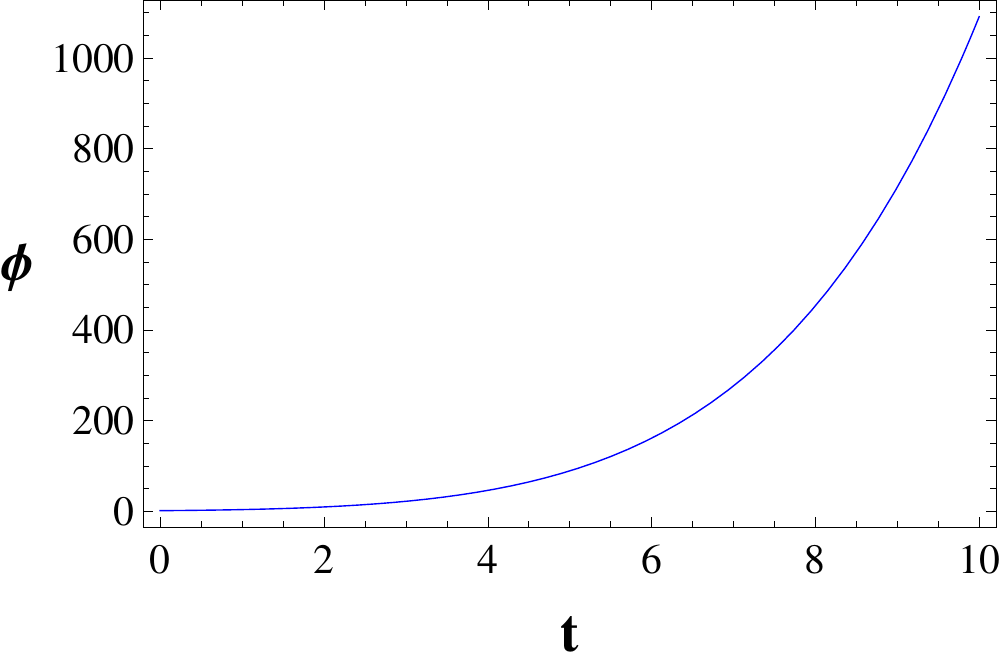}
\includegraphics[width=3in]{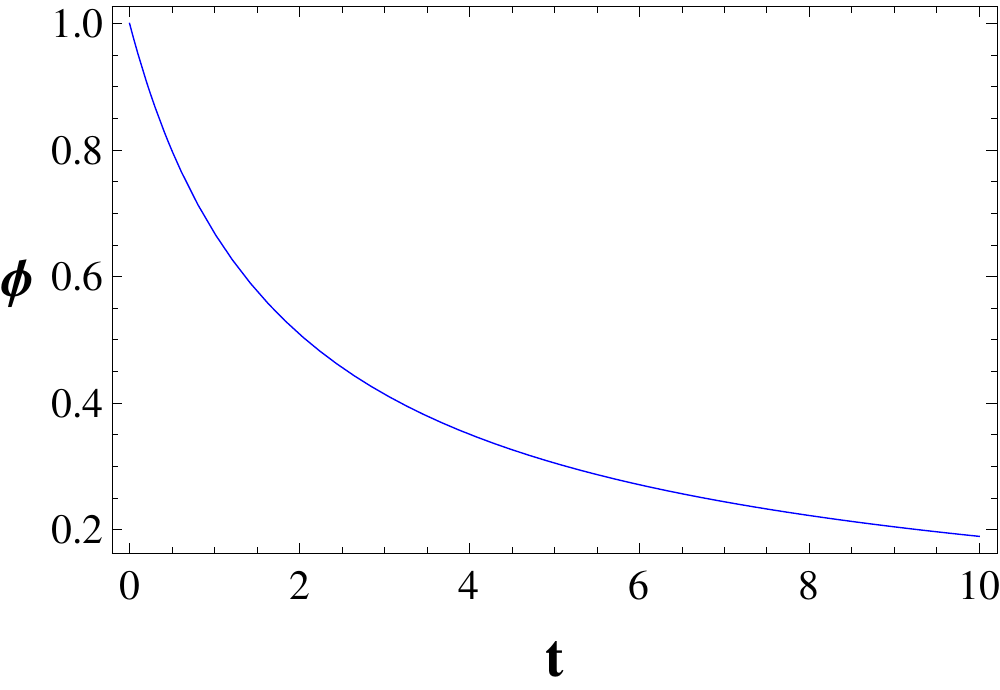}
\caption{{\footnotesize{{The time behavior of the
BD scalar field corresponding to slow (upper panel) and fast (lower panel) solutions.
We have taken $a_0=1=\phi_0$, $c_2=-1$, $\theta=0$ and $\omega=-1.2$.
Note that these curves, unlike their corresponding $a(t)$
curves [associated to both of the commutative and
noncommutative (with small values of the $\theta$) cases] coincide.
}}}}
\label{slow-fast}
\end{figure}

In the case where $\theta=0$, dependent on the value of
the BD coupling parameter, we get two types of solutions.
As these solutions might be essential for our discussion, let us obtain them
by the Hamiltonian formalism introduced in the previous section:
(i) when $g=-1/3$ (or $\omega=-4/3$), which corresponds to the
lower sign when $c_2>0$ and the upper sign when $c_2<0$,
the solutions describe the de Sitter--like space as
\begin{eqnarray}\label{de-sitter}
a(t)&=&a_{\rm i}\phi_{\rm i}^{-\frac{1}{3}}e^{mt} \hspace{3mm}{\rm and}
\hspace{5mm}\phi(t)=\phi_{\rm i}e^{-3mt},
\end{eqnarray}
where $\phi_{\rm i}$ is an integration constant,
and $m\equiv\frac{-|c_2|}{8a_{\rm i}^3}[-{\rm sgn}(c_2)\pm3]$,
ii) whereas, for $\omega\neq-4/3$, the solutions are in the power-law form as
\begin{eqnarray}\label{OT-solution-1}
a(t)&=&\tilde{a}_{\rm i}(t-t_{\rm ini})^{r_{\pm}},\\\nonumber
\phi(t)&=&\tilde{\phi}_{\rm i}(t-t_{\rm ini})^{s_{\pm}},
\end{eqnarray}
with
\begin{eqnarray}\label{OT-solution-2}\nonumber
\tilde{\phi}_{\rm i}\!\!\!&=&\!\!\!\Bigg\{\frac{\mid c_2\mid}{2a_{\rm i}^3\omega}
\left[{\rm sgn}(c_2)\mp\frac{(\omega+1)}{\xi}\right]\Bigg\}^{s_{\pm}},\\\nonumber
\tilde{a}_{\rm i}\!\!\!&=&\!\!\!a_{\rm i}\tilde{\phi}_{\rm i}^g
=a_{\rm i}\Bigg\{\frac{\mid c_2\mid}{2a_{\rm i}^3\omega}
\left[{\rm sgn}(c_2)\mp\frac{(\omega+1)}{\xi}\right]\Bigg\}^{r_{\pm}},
\end{eqnarray}
where $t_{\rm ini}$ is an integration constant and
the exponents $r_{\pm}$ and $s_{\pm}$ are given by
\begin{eqnarray}\label{r-s}
r_{\pm}&=&\frac{1}{3\omega+4}\left[\omega+1\pm {\rm sgn}(c_2)\xi\right],\\\nonumber
s_{\pm}&=&\frac{1\mp 3{\rm sgn}(c_2)\xi}{3\omega+4}.
\end{eqnarray}
Indeed, due to the general form of the above relations,
 the solutions~(\ref{OT-solution-1}) can be considered as a
 generalized version of the well-known O'Hanlon-Tupper
solution~\cite{o'hanlon-tupper-72-KE95-MW95,Faraoni.book} for a spatially flat FLRW universe.
Let us explain the role of the parameters present in the model.
In the special case where $c_2>0$ (or $c_2<0$), we obtain the solutions
corresponding to $(r_{+},s_{+})$ and $(r_{-},s_{-})$ known as
the fast and slow solutions, respectively~\cite{Faraoni.book}.
Such a designation can be related to the
behavior of the BD scalar field at $t\rightarrow0$
(for $\omega>-4/3$), such that the fast (slow) solution is
associated to the decreasing (increasing) BD scalar
field at early times.
In Fig.~\ref{slow-fast}, we have plotted these behaviors of
the BD scalar field for the fast and slow solutions\rlap.\footnote{We should note that, in
some situations, when showing plots in the same figure,
from rescaling the plots or manipulating the initial
conditions just for visual clarity, it may lead to incorrect physical
interpretations. Hence, the behaviors of these quantities are
plotted in separate figures; see, e.g.,  Fig.~\ref{slow-fast}.}

It has been shown~\cite{V91-GV92-TV92-GPR94} that when $\omega\neq-4/3
$, by redefining $\Phi\equiv-{\rm ln}(G\phi)$
(where $G$ is the gravitational constant), there are duality transformations as
\begin{eqnarray}\label{duality}
\alpha&\rightarrow&\left(\frac{3\omega+2}{3\omega+4}\right)
\alpha-2\left(\frac{\omega+1}{3\omega+4}\right)\Phi,\\\nonumber
\Phi&\rightarrow&-\left(\frac{6}{3\omega+4}\right)\alpha-
\left(\frac{3\omega+2}{3\omega+4}\right)\Phi,
\end{eqnarray}
 under which the slow and fast solutions are
 interchanged~\cite{L95-L96}, namely, $(r_{\pm},
 s_{\pm})\longleftrightarrow(r_{\mp},s_{\mp})$.
However, in our model for $\theta=0$ herein, from general relations~(\ref{r-s}),
without considering the duality transformations~(\ref{duality}),
we can see that the sign of the
integration constant $c_2$ is responsible for the
mentioned role, interchanging the lower-upper solutions.
More precisely, under interchanging $c_2>0\leftrightarrow c_2<0$,
 the parameters $c_1$, $g$, and $f$ transform
as $(c_1^{\pm}, f^{\pm}, g^{\pm})\longleftrightarrow (-c_1^{\mp},-f^{\mp},g^{\mp})$, and, consequently, we get $(r_{\pm},s_{\pm})\longleftrightarrow(r_{\mp},s_{\mp})$. By considering such a symmetry, a relevant counterpart
 between the solutions can be made, such that the number of
 %the investigations, for studying
 different cases to study are reduced by half.
 We also notice that, for the noncommutative case where $\theta\neq0$, as seen
from~(\ref{H}), the general duality transformations (if one can find them),
not only depend on the $f$, $g$, and the integration constants $c_1$ and
 $c_2$ but also may depend on the noncommutativity parameter.

In the rest of this section, we will investigate the results of a
few numerical solutions for the noncommutative case
which will be compared with the corresponding solutions of the commutative case.

\subsection{ Case I: The lower sign with $c_2>0$, $\omega<0$}
\begin{figure}
\centering\includegraphics[width=3.2in]{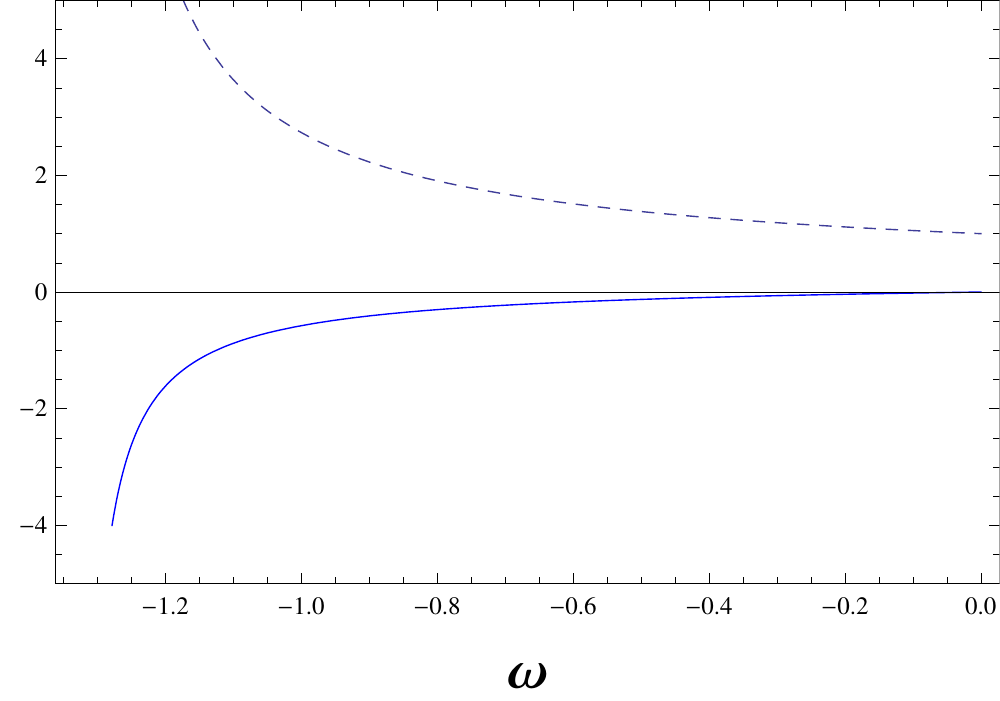}
\centering\includegraphics[width=3.2in]{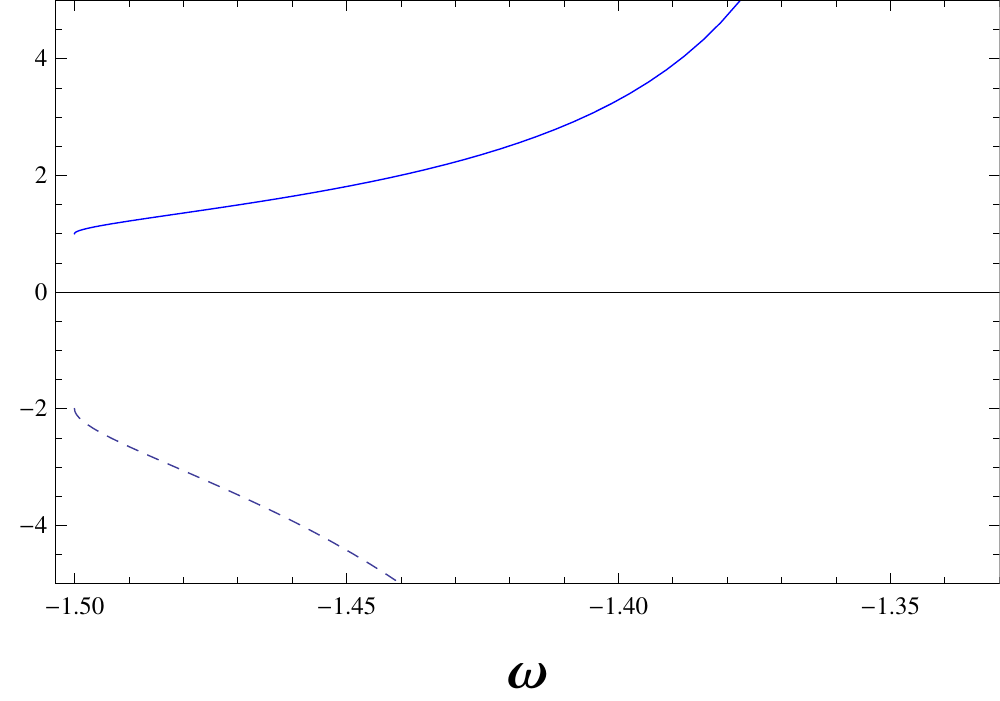}
 \caption{{\footnotesize{{The behavior of the exponents $r_{-}$ (solid curves) and
 $s_{-}$ (dashed curves) versus $\omega$ associated to the commutative case for the lower sign when $c_2>0$.
We have chosen the ranges $-4/3<\omega<0$ and $-3/2<\omega<-4/3$ for the upper and lower panels, respectively.}}}}
\label{phi.a-com}
\end{figure}

\indent
In the commutative case, for $c_2>0$, $\omega<0$, and the
lower sign, the time behavior of the scalar
field and scale factor depend on the values of
the $\omega$, in which, when $\omega$ is restricted
to $-3/2<\omega<-4/3$, the scalar field decreases while the scale factor always accelerates.
However, for $-4/3<\omega<0$,
we observe a different behavior for the scalar field and
scale factor, such that, for this case, the former increases but the latter
decreases. For this, in Fig.~\ref{phi.a-com}, according to the relations~(\ref{r-s}),
we have plotted the behaviors of the exponents
$r_{-}$ and $s_{-}$ versus $\omega$ in the range $-3/2<\omega<0$.
Hence, in order to have a simple comparison of
the commutative and noncommutative cases, perhaps it will be a good idea if we also
investigate these ranges of $\omega$ in separate parts for the
commutative and noncommutative cases. As the behavior of the quantities is sensitive to
the sign of the noncommutative parameter, we will
investigate various cases for positive and negative $\theta$.
%%%%%%%%%%%%%%%%%%%%%%%%%%%%%%%%%%%%%%%%%%%%%%%%%%%%%%%%%%%%%%%%%%%%%%%%%%55555
%%%%%%%%%%%%%%%%%%%%%%%%%%%%%%%%%%%%%%%%%%%%%%%%%%%%%%%%%%%%%%%%%%%%%%%%%%%%%%%%%555
%%%%%%%%%%%%%%%%%%%%%%%%%%%%%%%%%%%%%%%%%%%%%%%%%%%%%%%%%%%%%%%%%%%%%%%%%%%%5

\subsubsection{ {\bf Case Ia: $-3/2<\omega<-4/3$ and $\theta<0$}}

\begin{figure}
\includegraphics[width=3.2in]{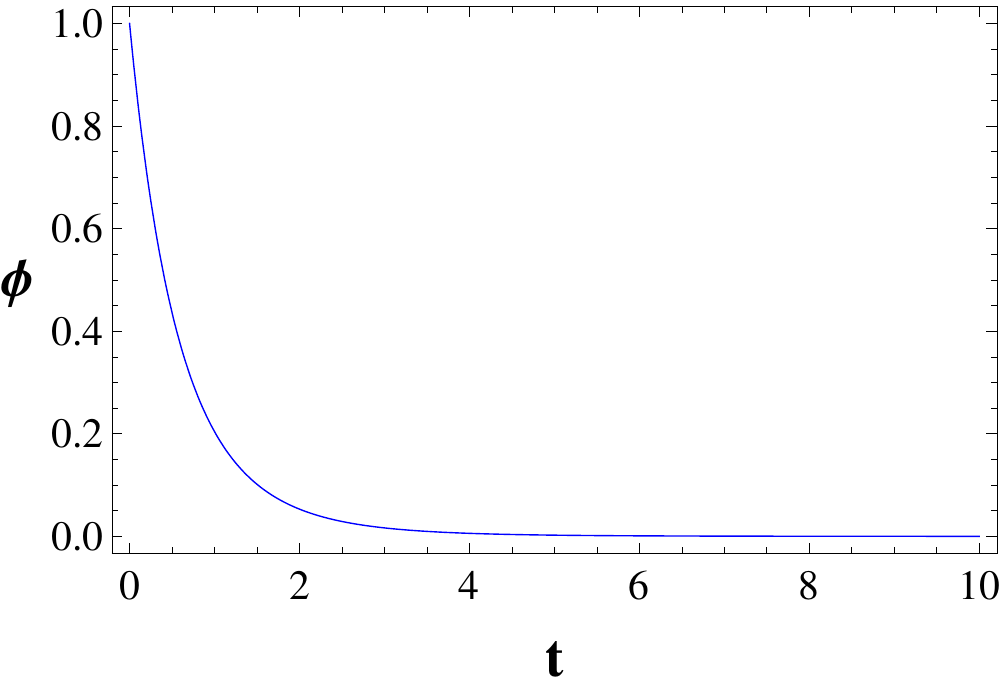}
\centering\includegraphics[width=3.2in]{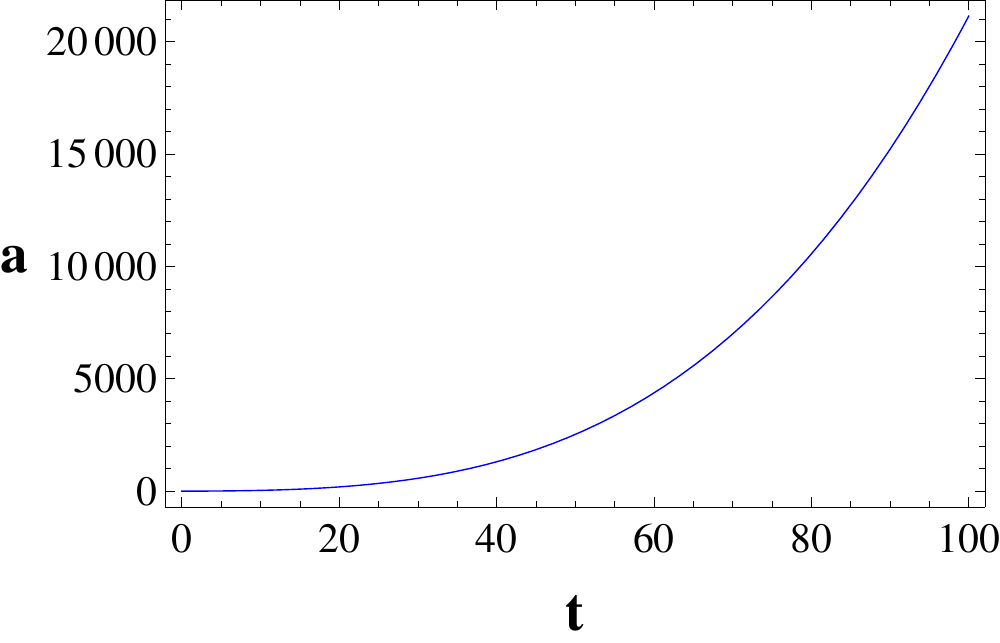}
\includegraphics[width=3.2in]{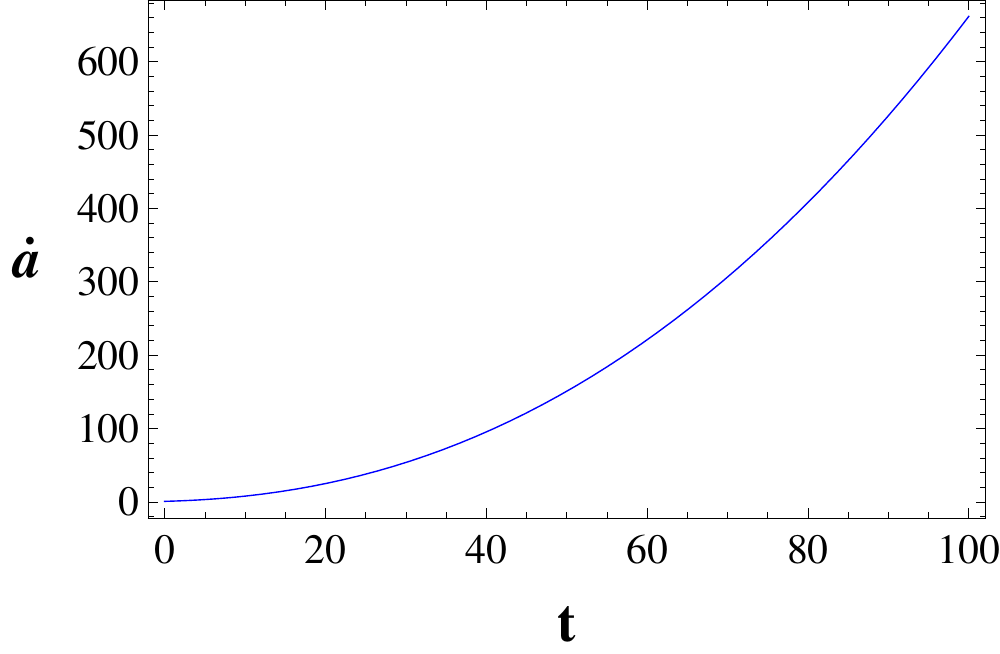}
\includegraphics[width=3.2in]{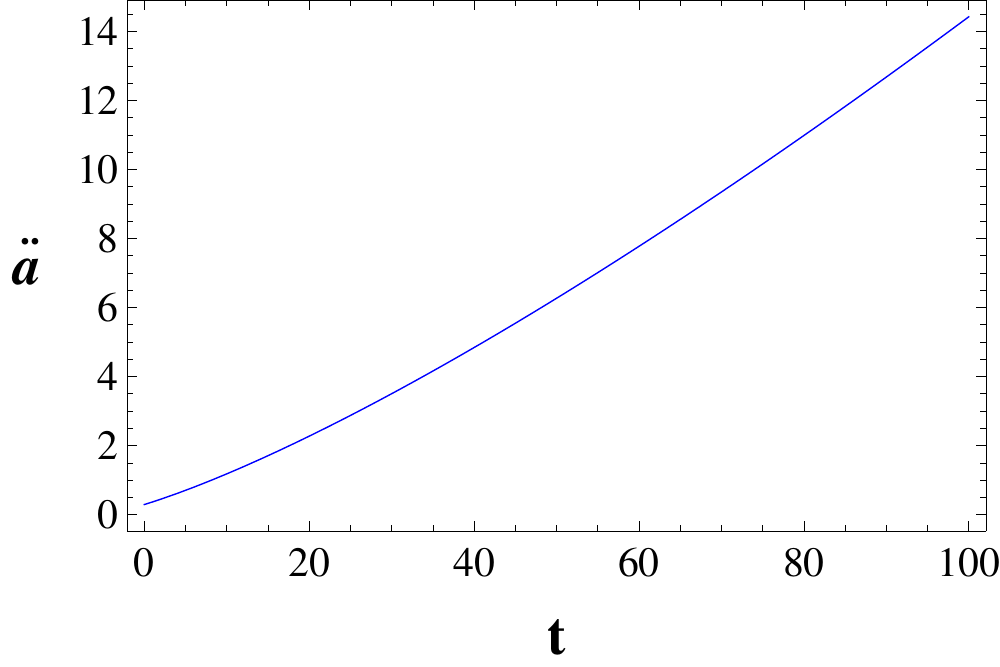}
 \caption{{\footnotesize{{The time behavior of the
BD scalar field, scale factor, its first and second derivatives associated to the
commutative case ($\theta=0$) for the lower
sign with $a_0=\phi_0=c_2=1$ and $\omega=-1.4$.
}}}}
\label{a-phi-case1}
\end{figure}

\begin{figure}
\centering\includegraphics[width=3.2in]{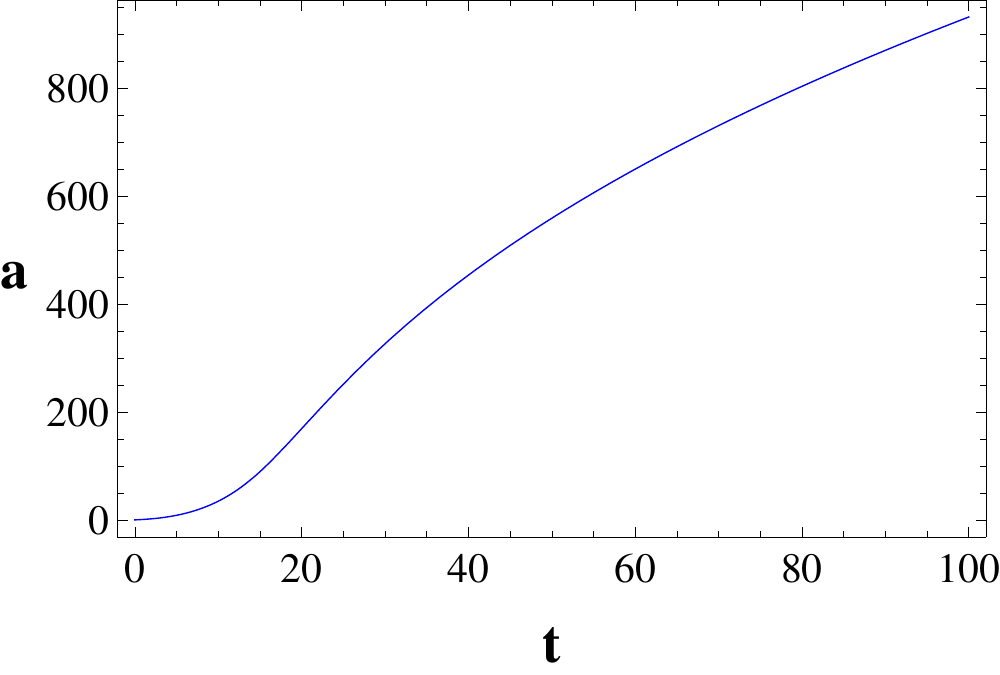}
\includegraphics[width=3.2in]{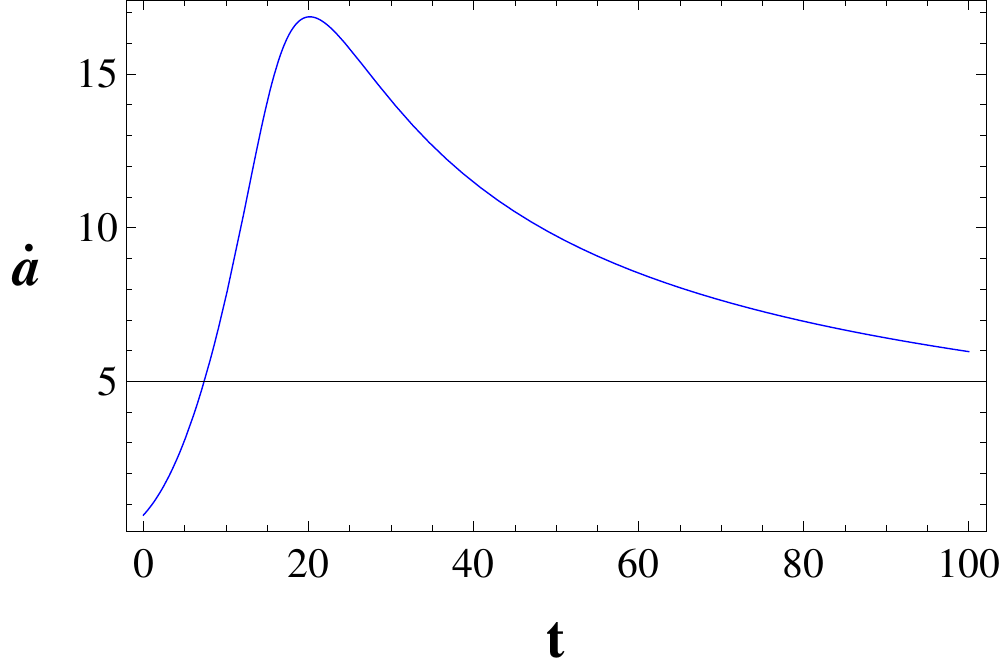}
\includegraphics[width=3.2in]{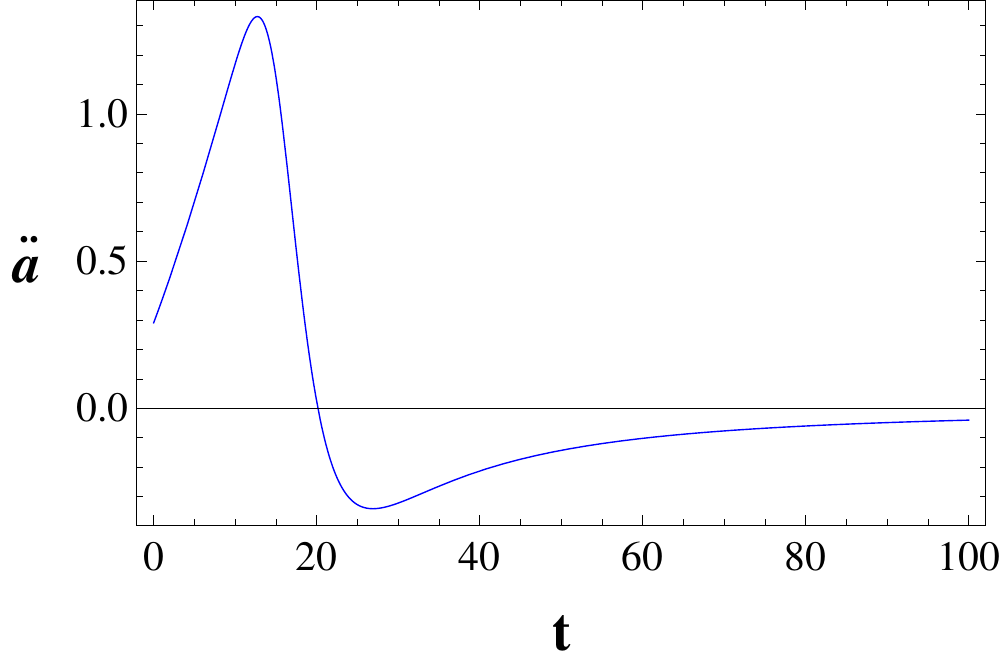}
 \caption{{\footnotesize{{The time behavior of the scale factor, its first and second
derivatives associated to the noncommutative case for the lower
sign. We have taken $\theta=-0.000001$. All the other
initial values are the same as they were in Fig.~\ref{a-phi-case1}.
Note that the $\phi(t)$ curves associated to the
commutative and noncommutative cases almost coincide.
}}}}
\label{a-phi-case1-NC}
\end{figure}

In order to compare the behavior of the quantities,
%associated to the commutative and noncommutative cases,
in Figs.~\ref{a-phi-case1}
and~\ref{a-phi-case1-NC}, we have plotted the time
behavior of the scalar field, scale factor, and its first
and second time derivatives for the commutative and
noncommutative cases, respectively.
In these plots, except for the noncommutative parameter,
we have chosen the same initial values for the variables:
very small negative values for the noncommutative
parameter and negative values for the BD parameter in the range $-3/2<\omega<-4/3$ and $a_0=\phi_0=c_2=1$.
We should remind that, in order to
probe the effects of the noncommutative parameter with more
clarity, Figs.~\ref{a-phi-case1} and~\ref{a-phi-case1-NC} have
been plotted separately for each case.

As Figs.~\ref{a-phi-case1} and~\ref{a-phi-case1-NC} show, the scalar field
decreases, and its behavior is almost the same for both of the
commutative and noncommutative cases (they almost coincide).
However,
%as seen from the figures~\ref{a-phi-case1} and~\ref{a-phi-case1-NC},
the behavior of the scale factor is different. That is, the scale factor
starts from a singular point at $t=0$ and increases
for both of the commutative and noncommutative cases, such that,
for the commutative case, we always have $\ddot{a}>0$, while for the noncommutative
case in the early times we have $\ddot{a}>0$, but
at the special point (hereafter, we call it ``point A''), it turns to be negative;
namely, after a very small time, the phase changes and we have a decelerating universe.
In the next sections, we will discuss further such an interesting behavior of the scale factor.
It is worthwhile to describe the evolution of the scale factor, scalar
field and their time derivatives for different values of the
three present parameters in this case. Namely, for
different values which have been chosen from the ranges $c_2>0$, $\theta<0$, and
$\omega<0$. The results indicate the following:
\begin{itemize}
  \item The larger the integration constant $c_2$, the shorter the
time of the accelerating phase (see Fig.~\ref{adotdot-diff-c2}). Namely, when we take a larger $c_2$,
the scale factor increases faster (with a larger speed and acceleration), and, consequently,
we get to the point A faster. In other words, by increasing
this integration constant, the $\ddot{a}$ curve is shifted
to the left simultaneously with a contraction of the amount of the time
associated to the accelerating (as well as decelerating) and an increase of $|\ddot{a}|$
for both the accelerating and decelerating phases.
Furthermore, as Fig.~\ref{adotdot-diff-c2} demonstrates, by taking a larger value for $c_2$,
the scalar field decreases faster.
Consequently, as the value of $c_2$ determines the time of the accelerating phase and
its corresponding scale factor value, it can, thus, be related to the number of $e$-folds for
an inflationary universe.
\begin{figure}
\centering\includegraphics[width=3.5in]{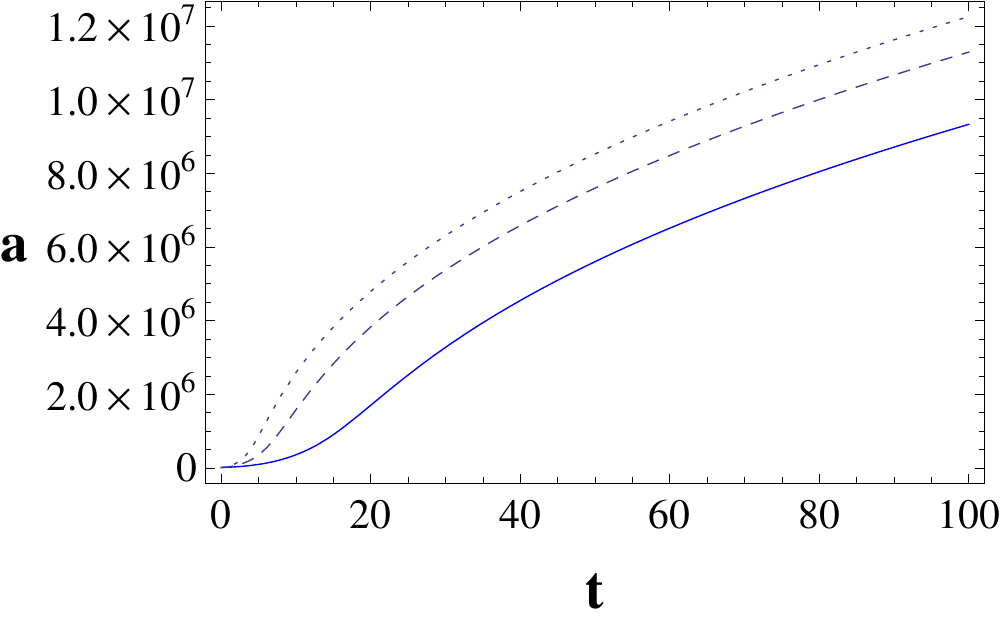}
\centering\includegraphics[width=3.5in]{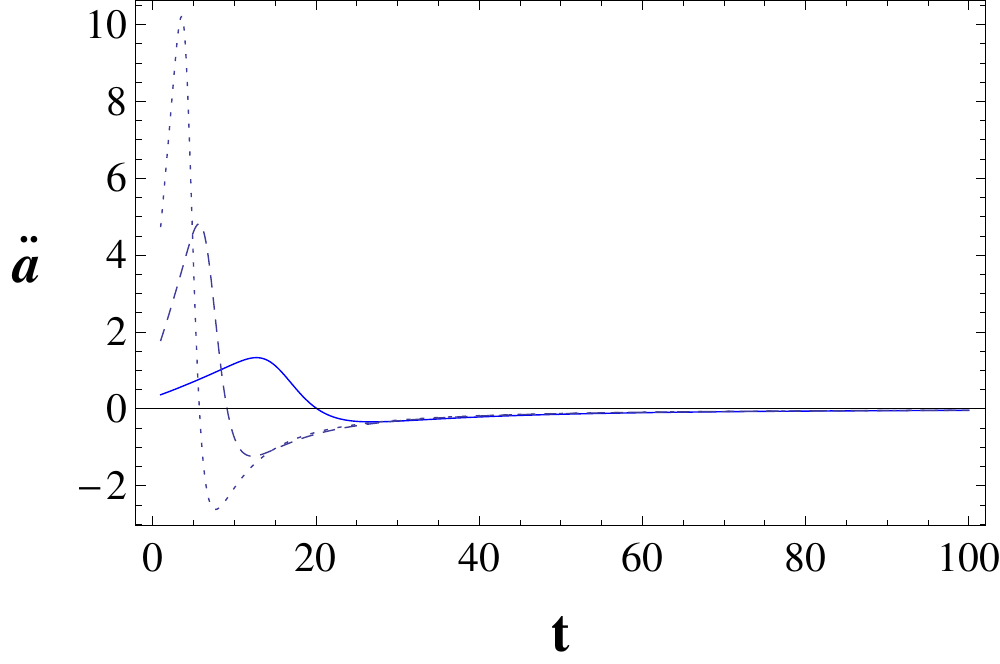}
\centering\includegraphics[width=3.5in]{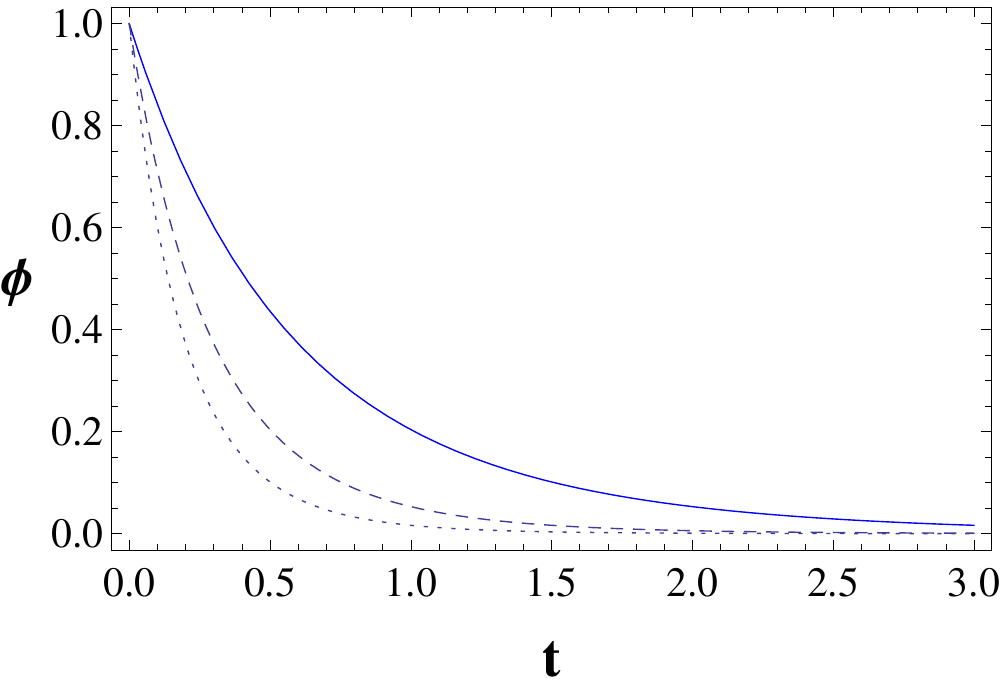}
 \caption{{\footnotesize{{The time behavior of the
$a$, $\ddot{a}$ and the $\phi$ for the noncommutative case with different $c_2$.
In this figure, we take three different values as $c_2=1$
(solid curve), $c_2=2$ (dashed curve) and $c_2=3$ (dotted curve).
The other parameters have the same initial values as in Fig.~\ref{a-phi-case1}}}}}
\label{adotdot-diff-c2}
\end{figure}
\item
In Fig.~\ref{adot-diff-theta}, we have plotted the
behavior of the scale factor and its
derivatives (with respect to the cosmic time) for different
values of the noncommutative parameter.
For different values of $\mid\!\theta\!\mid$, we cannot see
perceptible changes in the behavior of the scalar field.
However, the smaller the $\mid\!\theta\!\mid$, the larger
the slope of $a(t)$, namely $\dot{a}$.
More precisely, the speed of the expansion directly depends on
the $\mid\!\theta\!\mid$. More concretely, the
time behavior of $a$ (and, consequently, $\dot{a}$
and $\ddot{a}$), by taking different values
of the noncommutative parameter, changes, such that as
Figs.~\ref{adotdot-diff-c2} and~\ref{adot-diff-theta} show,
 it seems that $\mid\!\theta\!\mid$ plays the role of $c_2^{-1}$ (as the previous item shows) in the $\ddot{a}$ plots.
This claim is true for the amount of the interval
time of the accelerating and decelerating phases, but it does not
hold for the value of $\mid\!\ddot{a}\!\mid$, because in this case, the
larger the value of $\mid\!\theta\!\mid$, the smaller the value of $\mid\! \ddot{a}\!\mid$.
Hence, we should note that, as the numerical results show, it is not valid to argue that
when the value $\mid\!c_2\theta\!\mid$ remains constant, the behavior
of the scalar field, scale factor, and their time derivatives do not change.
This can be read from~(\ref{NC-a-phi}) and (\ref{NC-diff-phi});
namely, the extra role of $c_2$ in other parts of the differential
equations, for instance, the $f$ itself, depends on $c_2$ rather than $\theta$.
\begin{figure}
\centering\includegraphics[width=3.5in]{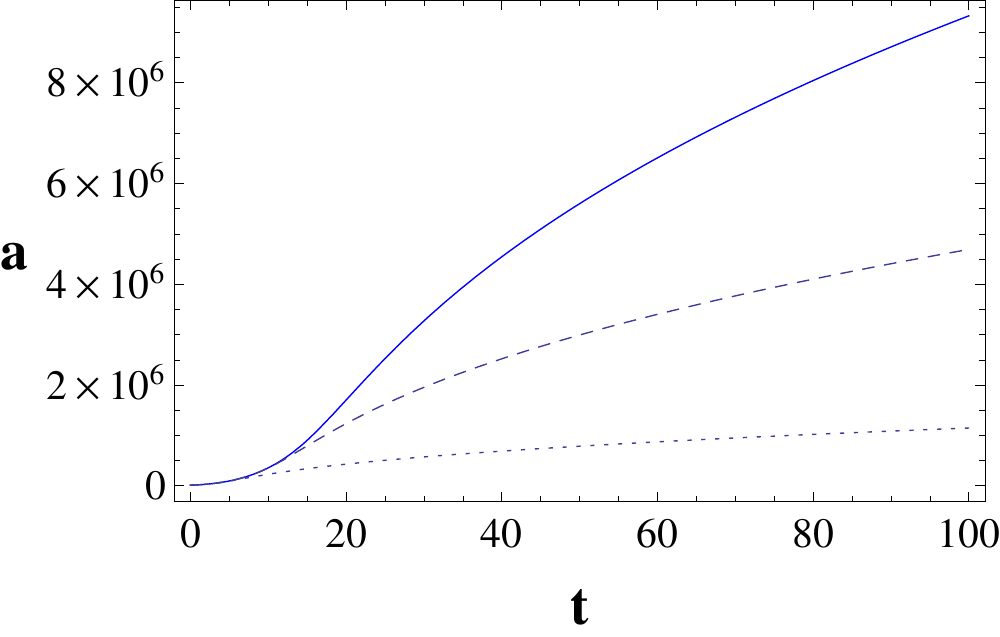}
\centering\includegraphics[width=3.5in]{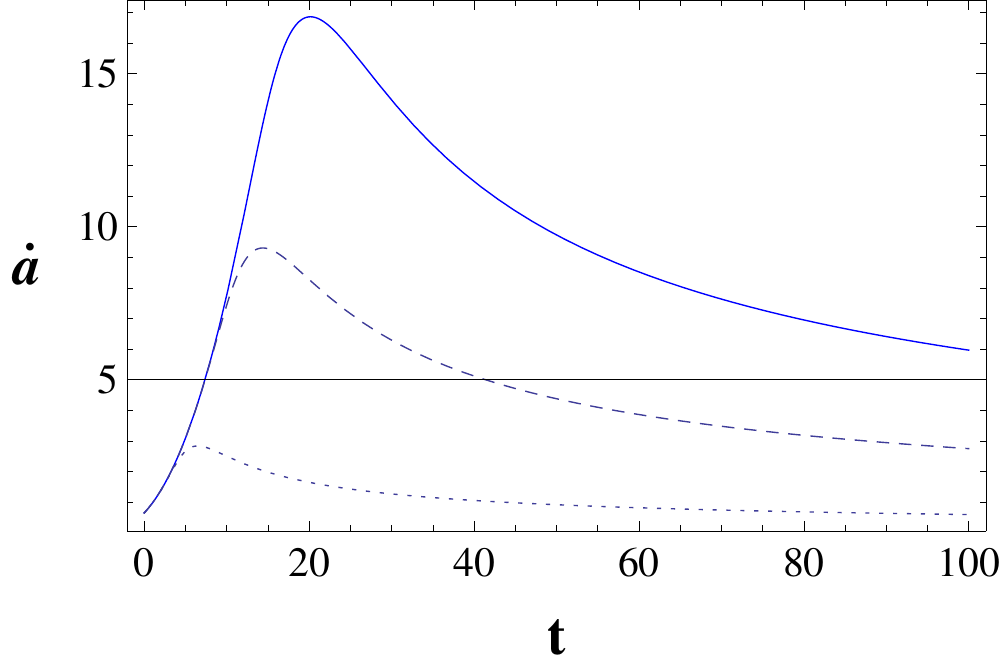}
\centering\includegraphics[width=3.5in]{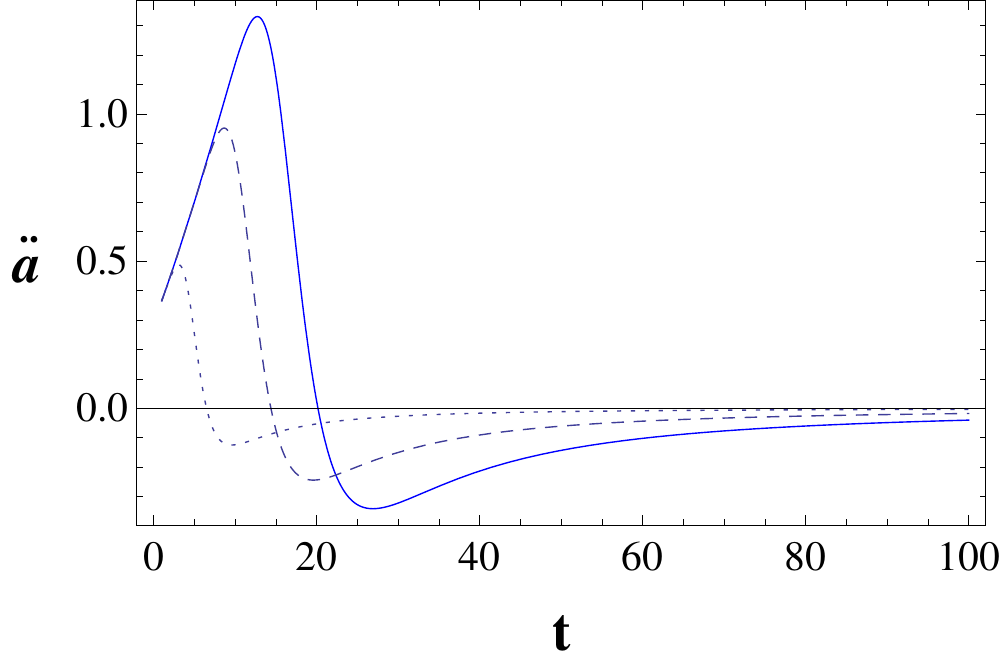}
 \caption{{\footnotesize{{The time behavior of the
$a$, $\dot{a}$ and $\ddot{a}$ for the noncommutative
case for different values of the noncommutative parameter.
Here, we take $c_2=1=a_0$, $\omega=-1.4$ for three
different values of the noncommutative parameter as $\theta=-0.000001$ (solid curve),
$\theta=-0.00001$ (dashed curve) and $\theta=-0.001$ (dotted curve).}}}}
\label{adot-diff-theta}
\end{figure}

\item
 For large values of the cosmic time, by assuming the same initial values for
 the parameters of the model (except $\theta$), the BD scalar field goes to zero for both
the commutative and noncommutative cases. However, the time behavior of the scale
factor is not the same for these cases. In the commutative case, the scale factor
always accelerates with a variable acceleration, such that $\ddot{a}$ never
takes a constant value. However, for the noncommutative case, $\ddot{a}$ vanishes, $\dot{a}$ takes
very small constant value, and, consequently, the Universe expands with a small constant speed.
Namely, in the noncommutative case, for late times, we get a zero acceleration epoch.
Such behavior for the scale factor can be interpreted as a direct consequence
of the existence of the noncommutative parameter.
This is almost
similar to the result obtained in~\cite{RFK11}, in which a constant deformation parameter is also included.
However, the difference is that, in our model, the speed of the
scale factor in late times is not exactly zero, but it approaches zero instead.
We should note that in~\cite{RFK11}, as the behavior of the quantities
were investigated, when $\omega\longrightarrow\infty$, such a
difference can be interpreted as a natural consequence of the models.
This effect of the noncommutative parameter
shows itself very far from the initial singularity, and
it has been suggested as a footprint of quantum
gravity in a coarse-grained explanation.
\end{itemize}
\subsubsection{ Case Ib: $-4/3<\omega<0$ and $\theta<0$:}
\begin{figure}
\centering\includegraphics[width=3.2in]{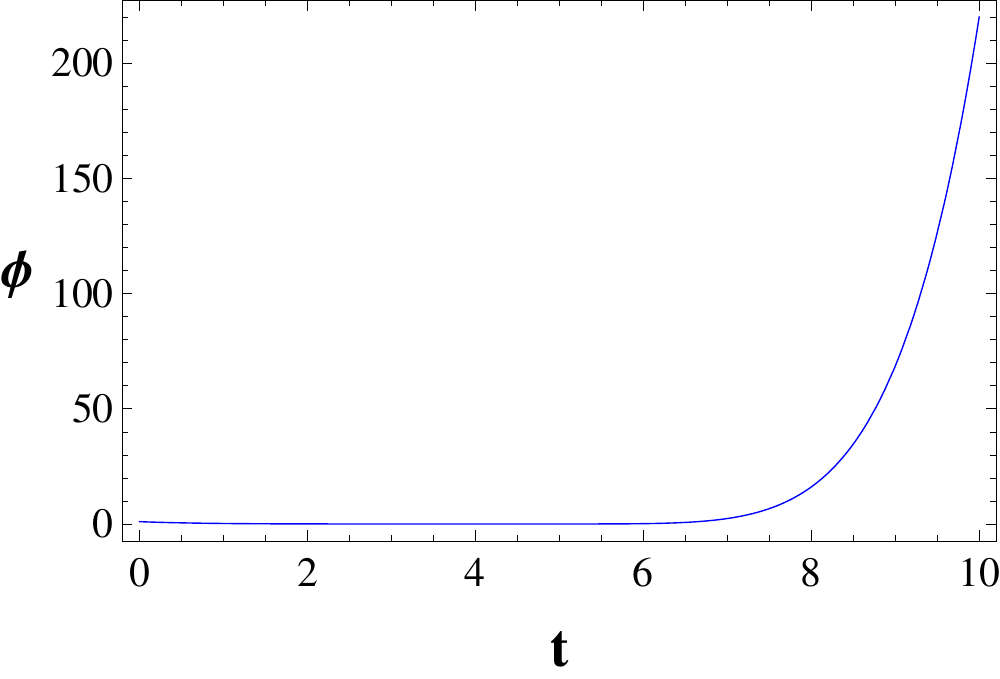}
\centering\includegraphics[width=3.2in]{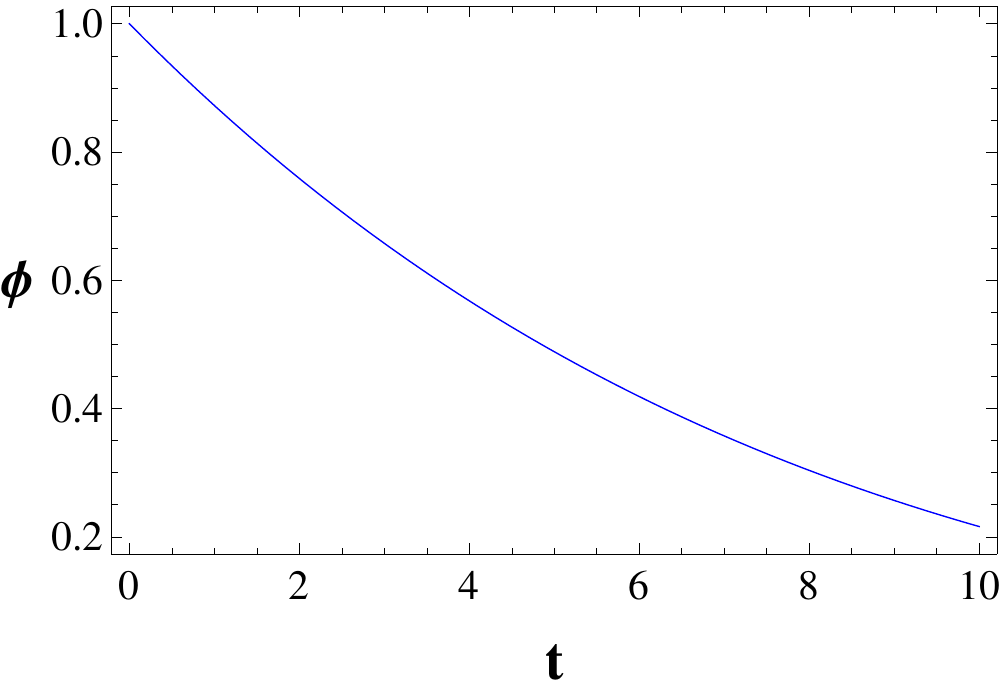}
 \caption{{\footnotesize{{The time behavior of the
BD scalar field for the commutative
case for different values for $c_2$ as $c_2=1$ (upper panel) and $c_2=0.1$ (lower panel).
The other initial values have been taken the same for both of the plots.
Here, $a_0=1$, $\omega=-1.2$ and $\theta=0$.}}}}
\label{phi-case-Ib}
\end{figure}

As mentioned, in this range of the BD coupling
parameter, for the commutative case, our solutions
%due to the appearance of the integration constant $c_2$,
are more general than the solutions obtained by O'Hanlon and Tupper.
More precisely, in the O'Hanlon-Tupper solutions, for the lower sign
with $-4/3<\omega<0$, the BD scalar field always increases while the scale factor
decreases. However, the behavior of these quantities in our model not only
depends on the values of $\omega$ but is also sensitive to the
values of the integration constant $c_2$, such that by
taking different values for $\omega$ and $c_2$, we
can obtain, in addition to the O'Hanlon-Tupper solutions other
different behaviors as obtained in case Ia.
For instance, in Fig.~\ref{phi-case-Ib},
we have plotted the behavior of the BD scalar field for two different values of $c_2$.
Note that the other initial values are the same for both of these figures.
Moreover, for the noncommutative case, we also observe that the behavior of these
quantities depends on, besides $c_2$ and $\omega$, the noncommutative parameter.
In short, for the noncommutative case, the obtained
solutions in the previous case (case Ia) can also be produced when we
take the range $-4/3<\omega<0$, although the initial values may be changed.

\subsection{Other cases}
We can also add a new case, i.e.,
the lower sign in which $c_2>0$, $\omega<0$, and $\theta>0$, instead
of negative values for the noncommutative parameter.
Also, we can analyze other cases similar to those categorized
in case I but instead with an upper sign (rather than a lower sign).
However, as all of the mentioned cases give different
results, which are not in the scope of this work, we will leave them.

 \section{Kinetic inflation}
 \indent
 \label{Kinetic inflation}
 In the previous section, we have shown that by introducing
 a noncommutative relation between the BD scalar field and
 the logarithm of the scale factor, not only does the
 scale factor accelerate in the early times, but also it can exit from
 the acceleration epoch and initiate a decelerating phase.
In other words, a suggestion on how to solve the graceful exit problem.
 However, these features alone do not guarantee
 an appropriate setting for the resolution of
 the problems with the standard cosmology.

 One of the well-known shortcomings with the standard cosmology is the horizon problem.
 Namely, there are plenty of regions in the large
 volume of today's Universe, which were not causally connected at early times.
 More precisely, the size of the
 presently\footnote{A subscript $0$ stands for the present epoch.}
 observed Universe at some earlier time $t$ (at least as
 early as $t\sim1 sec$), $d_{_{\rm U}}(t)\sim a(t)/(H_0a_0)$, is much larger than a distance
 which a photon traveling by\footnote{The primed variables are evaluated at time $t'$.
 We should note that the quantity introduced here
 as $d^{^{\rm Hor}}$ is, indeed, the radius of the optical horizon defined
 for the FLRW space in which $t_i=t_{\rm recombination}$~\cite{F11}.
 While, the radius of the particle horizon
 at time $t$ usually is defined as a radius of a sphere whose center is located
 at the same point where the comoving observer localized, and it encompasses
 all particle signals have been reached from the time of the
 big bang, (i.e., $t=0$, instead of $t_{\rm recombination}$)
 until $t$.} $t$, $d^{^{\rm Hor}}(t)=a(t)\int_{t_i}^{t}dt'/a'$~[28].
  %More precisely, the size of the Universe that is observed
 %presently\footnote{A subscript $0$ stands for the present epoch.}at a
 %definite time $t$, $d_{_{\rm U}}\sim a(t)H_0^{-1}/a_0$ is much larger than a distance
 %which a photon traveling by that
 %time\rlap,\footnote{The primed variables are evaluated at time $t'$.
 %We should note that the quantity introduced here
 %as $d^{^{\rm Hor}}$ is indeed the radius of the {\it optical horizon} defined
 %for the FLRW space in which $t_i=t_{\rm recombination}$~\cite{F11}.
 %Whilst, the radius of the {\it particle horizon}
 %at time $t$, usually, is defined as a radius of a sphere whose center is located
 %at the same point where the comoving observer localized, and it encompasses
 %all particle signals have been reached from the time of the
 %big bang, (i.e., $t=0$, instead of $t_{\rm recombination}$) until $t$.}
 %$d^{^{\rm Hor}}=a(t)\int_{t_i}^{t}dt'/a'$~\cite{HTW94}.
Let us first check the nominal condition
for the acceleration associated to the inflation~\cite{Lev95}, namely,
 \begin{eqnarray}\label{NC-inflation-1}
d^{^{\rm Hor}}>H^{-1}.
\end{eqnarray}
 Then, in the rest of this section, we will investigate the condition for sufficient inflation.

From Eqs.~(\ref{phi-dot}) and (\ref{H}), we get
\begin{eqnarray}\label{NC-inflation-2}
 \left(H+\frac{\dot{\phi}}{2\phi}\right)^2=
 \left[\left(h+\frac{1}{2}\right)\left(\frac{\dot{\phi}}{\phi}\right)\right]^2,
\end{eqnarray}
which gives
\begin{eqnarray}\label{NC-inflation-3}
 \frac{d{\rm ln}(a^2\phi)}{dt}=\pm(2h+1)\left(\frac{\dot{\phi}}{\phi}\right).
\end{eqnarray}
This equation shows that the horizon distance
can be related to $\phi$, $\omega$, the scale factor, and the noncommutatvity parameter.
Applying (\ref{phi-dot}) and integrating over $dt$, we obtain
%the following relation for the particle horizon
\begin{eqnarray}\label{NC-inflation-4}
 d^{^{\rm Hor}}=\frac{a^3\phi}{|f|}-\frac{2c_2}{2g+1} d^{^{\rm NC}}
\end{eqnarray}
up to a constant of integration. In Eq.~(\ref{NC-inflation-4}), we have introduced
the new distance $d^{^{\rm NC}}$ as
\begin{eqnarray}\label{d-NC}
 d^{^{\rm NC}}\equiv\frac{\theta a}{c_2}\int \frac{P'_{\phi'}dt'}{a'},
\end{eqnarray}
in which the integrand not only depends on the
inverse of the scale factor (similar to the one defined for optical horizon)
but also depends on the nonocommutativity parameter and the
conjugate momentum of the BD scalar field. The factor $c_2$ is multiplied in the
denominator of relation~(\ref{d-NC}) to make the
dimension of $d^{^{\rm NC}}$ the same as $d^{^{\rm Hor}}$.
(Note that we have found a relation between the
BD scalar field and its momentum conjugate as $P_\phi=c_2/\phi$.)
We expect that this new term can add a positive
value to the $d^{^{\rm Hor}}$ to properly assist in satisfying the
requirement associated to the horizon problem.
In order to compare, we rewrite Eq.~(\ref{H}) by the aid of (\ref{phi-dot}) as
\begin{eqnarray}\label{NC-inflation-5}
 H=\frac{|f|}{\xi}\frac{h}{a^3\phi}.
\end{eqnarray}
Using (\ref{NC-inflation-4}) and (\ref{NC-inflation-5}) in the
nominal condition (\ref{NC-inflation-1})
gives
\begin{eqnarray}\label{Nominal-con}
D^{^{\rm NC}}\equiv d^{^{\rm Hor}}\!\!-H^{-1}=
\frac{\phi a^3}{|f|}\left(\!\!1+\frac{\xi}{h}\right)-\frac{2d^{^{\rm NC}}}{2g+1}>0.
\end{eqnarray}
Obviously, in the limit $\theta\longrightarrow0$,
$d^{^{\rm NC}}$ goes to zero as well, and, thus, the relation
associated to the horizon distance of the commutative case is recovered.
%Clearly, this requirement reduces to its corresponding
%commutative case when the noncommutative parameter goes to zero.
Further, in the mentioned limit,
the resulted relation is the same as one obtained in Ref.~\cite{Lev95}
(by assuming a constant BD coupling parameter in the mentioned paper).
Therefore, in the commutative case where $\theta=0$, the
only acceptable result is $0<\xi<1$ ($-3/2<\omega<0$), which is obtained by
choosing either the upper sign for $c_2>0$ or the lower sign for $c_2<0$.

\begin{figure}
\centering\includegraphics[width=3.2in]{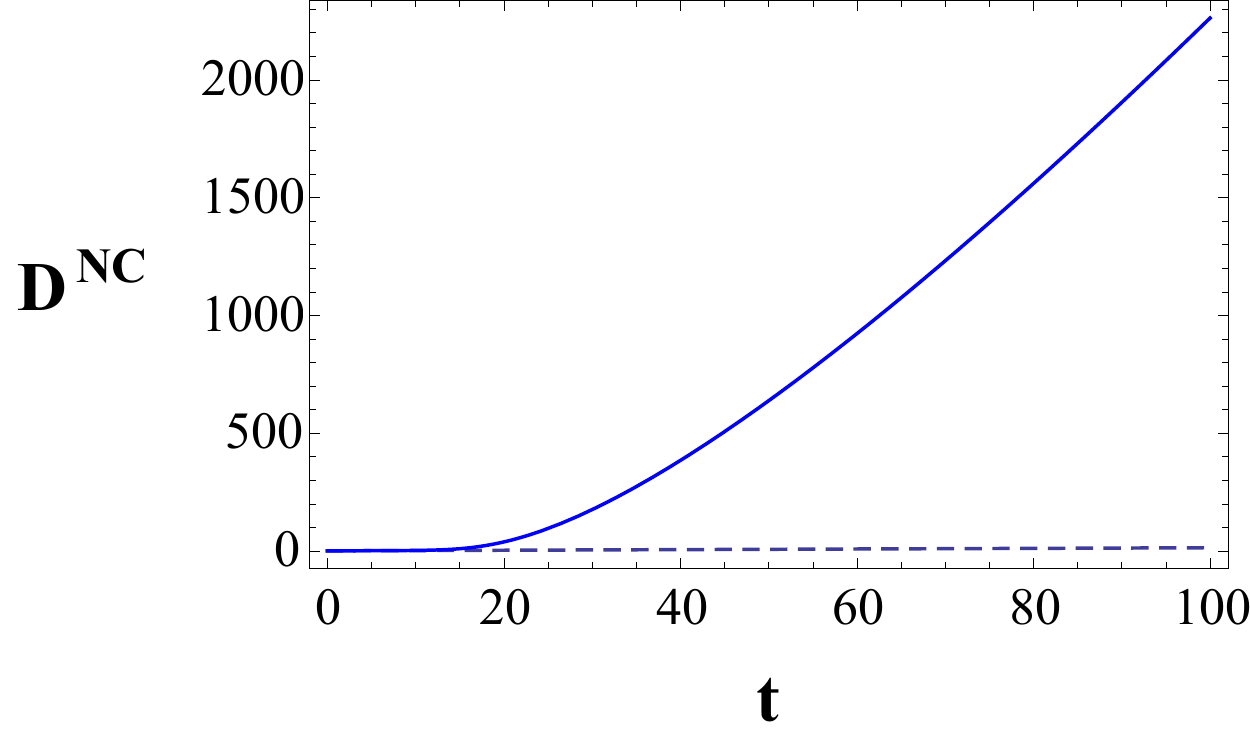}
 \caption{{\footnotesize{{The time behavior of the
 $D^{^{\rm NC}}$, the quantity which defined as in (\ref{Nominal-con}). The solid and dashed
curves are associated to the noncommutative and commutative cases, respectively.
 This figure is prepared as an example to indicate that the nominal condition~(\ref{NC-inflation-1})
 for the noncommutative solutions (specially for the case Ia) can be easily satisfied.
The initial values are $a_0=1=c_2$, $\omega=-1.4$, $\theta=-0.000001$
(solid curve) and $\theta=0$ (dashed curve).}}}}
\label{NR}
\end{figure}

\begin{figure}
\centering\includegraphics[width=3.2in]{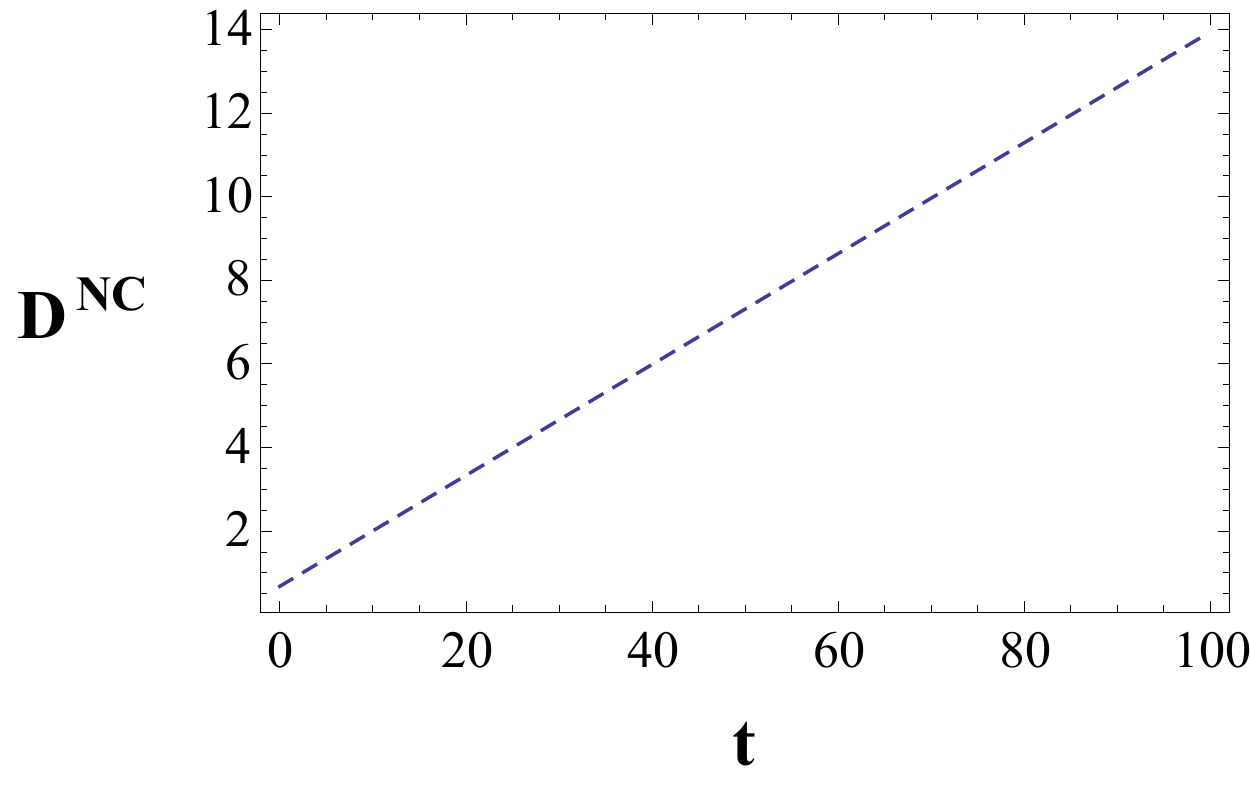}
 \caption{{\footnotesize{{The time behavior of the $D^{^{\rm NC}}$ for the
 commutative case with the same initial values as in Fig.~\ref{NR}.}}}}
\label{NR-Com}
\end{figure}

For the general noncommutative case, we
should note that, in addition to the sign of $c_2$, the allowed
values of $\omega$ as well as the
behavior of the BD scalar, the noncommutative parameter has a substantial role in determining
whether this constraint is satisfied or not.
Obviously, as the numerical results of the previous
section show (see, especially, Figs.~\ref{adotdot-diff-c2} and \ref{adot-diff-theta}
and their analysis), due to the presence of the noncommutative parameter and
the extra terms associated to it, satisfying the constraints for the
noncommutative case is easier than its corresponding commutative case.
For instance, in Fig.~\ref{NR}, for case Ia, $D^{^{\rm NC}}$ has been plotted
against cosmic time. Therefore, we observe that
the constraint~(\ref{Nominal-con}) for
the noncommutative case can be easily satisfied at all times.
To be more clear, we also have plotted the time
behavior of $D^{^{\rm NC}}$ for the commutative
case ($\theta=0$) separately in Fig.~\ref{NR-Com}.

In what follows, we intend to probe the
condition for sufficient inflation\rlap,\footnote{It has been
claimed~\cite{LF94} that the constraint~(\ref{suf-hor-3}) is only valid for the
power-law scale factor of the Universe. As in our
model, we assume that
the warp factor can be expanded such that, because of the
smallness of the noncommutative parameter, we can take only up to the linear term.
Then, the mentioned causality condition needed to overcome the
horizon problem holds for our model.} which is given by~\cite{Lev95-2}
\begin{eqnarray}\label{suf-hor-3}
 \frac{{d}^{^{\rm Hor}}_\star}{a_\star}>\frac{1}{H_0a_0},
\end{eqnarray}
where the lhs of the above inequality
stands for a comoving size of a causally connected region at a
specific earlier time $t_\star$.
 By including a nonzero integration constant, relation~(\ref{NC-inflation-3}) for
 the specific time $t_\star$ gives
 \begin{eqnarray}\label{suf-hor-con-2}
 d^{^{\rm Hor}}_\star=\frac{a^3\phi(1-\delta)}
 {f {\rm sgn}(c_2)}-\frac{2}{2g+1} d^{^{\rm NC}}\Biggr|_\star,
\end{eqnarray}
where the integration constant, which was removed in relation~(\ref{NC-inflation-4}), has now
been included in $\delta\equiv\frac{a_i^2\phi_i}{a^2\phi}$ where the subscript $i$ stands for
initial values. Note that, as the BD scalar field takes positive values, we always have $\delta\geq0$.
%By substituting $f$ from (\ref{phi-dot}) into~(\ref{suf-hor-con-2}), we get
%\begin{eqnarray}\label{suf-hor-con-3}
% d^{^{\rm Hor}}=\pm\left(\frac{\phi}{\dot{\phi}}\right)\left[\frac{1-\delta}{2h+1}\right].
%\end{eqnarray}

The Hubble constant at present time, $H_0$, can be expressed
in terms of the value of the Planck mass today, $M_0$, and $T_0$ as~\cite{Lev95-2}
\begin{equation}\label{h0}
H_0=\sqrt{\hat{\alpha}_0}\frac{T_0^2}{M_0},
\end{equation}
where $\hat{\alpha}_0=\gamma(t_0)\eta_0=(8\pi/3)(\pi^2/30)\bar{g}(t_0)\eta_0$, in which $\eta_0$ stands
for the ratio today of the energy density in matter to that in radiation.
 In order to see whether or not the above condition is satisfied by
 the solutions herein, we would like to employ the assumptions of the Ref.~\cite{Lev95-2}:
 (i) the time $t_{\rm end}$ is allocated to the end of inflation in which the entropy is produced;
 (ii) since the time $t_{\rm end}$, the Universe
 has evolved adiabatically such that we can assume
 $a_{\rm end}T_{\rm end}=a_0T_0$.
  Employing relation~(\ref{h0}) and assumption (ii) in (\ref{suf-hor-3}) gives
 \begin{eqnarray}\label{suf-hor-4}
 \frac{a_\star}{a_{\rm end}}\gtrsim\left(\frac{M_0}{\sqrt{\bar{\alpha}_0}T_0}\right)
 \frac{1}{\mid\!\!d^{^{\rm Hor}}_\star\!\!\mid T_{\rm end}}.
 \end{eqnarray}
 %where ${d}_{_{\rm Hor\star}}$ evaluated by relation (\ref{suf-hor-con-3}).
 In order to proceed, we
 consider a simple conjecture for the heating mechanism.
 Let us use the following relation between $T_{\rm end}$ and
 the net available kinetic energy $E_{\rm end}$ as
 \begin{eqnarray}\label{TE}
 T_{\rm end}=\epsilon E_{\rm end},
 \end{eqnarray}
 where $\epsilon$ denotes the efficiency of the
 system where the kinetic energy density is converted to entropy~\cite{Lev95-2}.
 The kinetic energy density for our model is only given by
 the energy density of the BD scalar field in unit volume. Namely, we have
  $E_{\rm end}=\rho^{(\phi)}_{\rm end}(4\pi/3)a^3_{\rm end}$, where by substituting
the energy density associated to the BD scalar field from relation~(\ref{rho-phi}), we obtain
%\begin{eqnarray}\label{p-ro-1}
 %\rho_\phi=3H^2\phi, \hspace{5mm} {\rm and}
 % \hspace{5mm} p_\phi=-\left(2\frac{\ddot{a}}{a}+H^2\right)\phi
%\end{eqnarray}
%\begin{eqnarray}\label{p-ro-2}
 %\rho_\phi=3h^2\left(\frac{\dot{\phi}^2}{\phi}\right),
 % \end{eqnarray}
 % p_\phi=\left[3h^2+2h+\frac{2c_2\theta}{\phi}\right]\left(\frac{\dot{\phi}^2}{\phi}\right)
%Thus, by using \ref{phi-dot}, we get
\begin{eqnarray}\label{T-end}
T_{\rm end}=\frac{4\pi\epsilon f^2}{\xi^2} \frac{h^2_{\rm end}}{a^3_{\rm end}\phi_{\rm end}}.
  \end{eqnarray}
  Substituting $d^{^{\rm Hor}}_\star$ from~(\ref{suf-hor-con-2}) and $T_{\rm end}$
  from (\ref{T-end}) into (\ref{suf-hor-4}), we get
\begin{eqnarray}\label{suf-hor-p}
 \frac{a_\star^2\phi_\star}{a_{\rm end}^2\phi_{\rm end}}&\gtrsim&
 \left(\frac{M_0}{\sqrt{\bar{\alpha}_0}T_0}\right)(1-\delta_\star)^{-1}\xi^2\\\nonumber
 &\times&\frac{1}{4\pi\epsilon \mid f\mid (g+\frac{c_2\theta}{\phi})^2
 [1-\frac{2d^{^{\rm NC}}_\star}{(2g+1)\mid f\mid(1-\delta_\star)a_\star^3\phi_\star}]}.
 \end{eqnarray}
 The constraint~(\ref{suf-hor-p}) is the modified
 (noncommutative) version of the one obtained in
 the BD theory in \cite{Lev95-2}.

 Let us first review the obstacles of the commutative model regarding~(\ref{suf-hor-3}) and
 then turn to solve the problems by applying the noncommutative model.
 In \cite{Lev95-2}, $(1-\delta_\star)\sim1$, $f\sim1$,
 $M_0=1.2\times10^{19}{\rm GeV}$, $T_0=2.3\times10^{-13}{\rm GeV}$, and assigning two
 different values for $\omega$, the model
 was examined. In one case where $\omega\sim0$, satisfying~(4.8) leads us to take the D branch.
 %To satisfy the requirement for sufficient inflation, it
% has been assumed that $\omega\sim0$, in which to
 %satisfy~(\ref{suf-hor-3}), it is needed to take the D branch.
 Let us be more precise.
  By substituting the above values in the inequality for the commutative case, the constraint
  $(a_\star^2\phi_\star)/(a_{\rm end}^2\phi_{\rm end})\gtrsim 10^{30}/\epsilon$ follows.
  This condition, even with $\epsilon\sim1$, implies that the
  quantity $a^2\phi$ decreases with the cosmic time.
  Such a result demands that we must take the D branch of the solutions.
  However, to have an expanding universe, we must have $a_{\rm end}>a_\star$.
  By using this requirement in (\ref{suf-hor-p}) for the
  commutative case, the minimum change in the dynamical Planck
  mass will be~
   $m_{\rm Pl}(t_{\rm end})/m_{\rm Pl}(t_\star)\gtrsim 10^{-15}\epsilon^{1/2}$,
   which implies that the Planck mass
   must decrease during inflation. However, as was argued
   by Levin~\cite{Lev95-2}, it is not enough that
   such a requirement is satisfied, and a branch change must be induced.

\begin{figure}
\centering\includegraphics[width=3.2in]{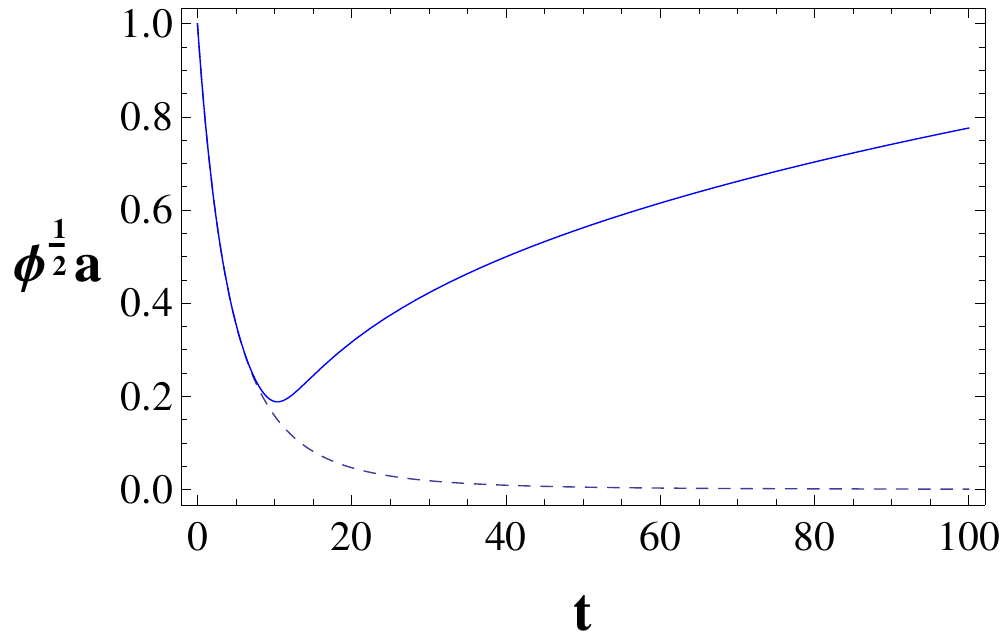}
 \caption{{\footnotesize{{The time behavior of the $a\phi^{1/2}$ for the
 commutative case (dashed curve) and noncommutative case (solid curve).
 We have taken $c_2=1=a_0$, $\omega=-1.36$ and $\theta=-0.000001$ for the noncommutative case.}}}}
\label{D-branch}
\end{figure}

   Let us further discuss the assistance of a few plots from our model
   for the lower sign with $c_2>0$ (D branch) and then compare it with the
   noncommutative case. Equation~(\ref{NC-inflation-3}),
   by using~(\ref{H}) and (\ref{g}) for the lower sign, can be rewritten as
   \begin{eqnarray}\label{D-b}
 \frac{d{\rm ln}(a\phi^{\frac{1}{2}})}{dt}=-\frac{1}{2}\Big[{\rm sgn}(c_2)\xi+\frac{2c_2\theta}{\phi}\Big]
 \left(\frac{|f|}{\xi a^3\phi}\right).
\end{eqnarray}
In the commutative case with $c_2>0$, we get
 \begin{eqnarray}\label{D-b2}
 \frac{d{\rm ln}(a\phi^{\frac{1}{2}})}{dt}=-\frac{|f|}{2a^3\phi},
\end{eqnarray}
 which indicates that the quantity $a\phi^{1/2}$ always decreases with the cosmic time.
 In Fig.~\ref{D-branch}, such behavior has been shown for the lower sign (see the dashed curve).
  Because a variation in the strength of gravity
 for today's Universe has not been observed, this behavior is consequently not acceptable:
  namely, the BD scalar field must take almost constant values today. On the other hand, as
  we have an expanding universe, the quantity $a\phi^{1/2}$ has to increase.
 While for a general noncommutative case, due to the complicated dependence of
 the rhs of~(\ref{D-b}) to $c_2$, $\omega$, $\theta$, as well as
 the BD scalar field, we cannot analytically draw the time behavior of $a\phi^{1/2}$.
 However, fortunately, for $c_2>0$, $\theta<0$, and lower case, i.e., the D branch, we have
 shown numerically that at the early times, $a\phi^{1/2}$ behaves the
 same as its corresponding in the commutative case; namely, it
 decreases with time. However, after reaching a nonzero minimum, it starts to increase.
 For instance, in~Fig.~\ref{D-branch}, the time behavior of
  $a\phi^{1/2}$ for the lower sign associated to the noncommutative case has been plotted.

  We also should remind that in the commutative case, even with taking a variable
  BD coupling parameter, the D branch cannot give today's expanding
  Universe. Hence, such a result is not consistent
 with the present accelerating Universe.
 % In the second case, the
 %$\omega$ has been assumed near $-3/2$ which leads to a
% mechanism in which the X-branch can meet the requirement~(\ref{suf-hor-3}).
 %However, it was stressed that, for this case, some fine tuning should be included~\cite{Lev95-2}.
%Note that $f$ and $g$ have been constructed from the
%constant parameters, so their values are the same for the
%times associated to $t_\star$ and $t_{\rm end}$. With the aid of this point, taking
%$c_2$ and $\xi$ as of order unity and $\theta<<<1$

\section{conclusions}
\label{Conclusions}
In this paper, we have introduced a noncommutative version of the BD theory.
More precisely, a modified Poisson algebra among
minisuperspace variables (the logarithm of the scale factor
and the BD scalar field) has been used.
Such an ansatz bears much resemblance to the assumptions taken in
noncommutative quantum cosmology~\cite{GOR02,BP04-PM05-AAOSS07-GSS07}
as well as a few classical noncommutative
cosmological models in theories alternative to general relativity
with a minimally~\cite{GSS11,RZMM14} (or a nonminimally~\cite{RFK11})
coupled scalar field to the geometry.

We have investigated the BD cosmological equations of motion in the comoving gauge.
The general Hamiltonian equations indicate that when the noncommutative
parameter tends to zero, all the equations reduced to their corresponding
counterparts in the standard commutative case.

We have focused on the case in which there is neither a scalar potential
nor a cosmological constant. Furthermore, we have assumed that the Lagrangian
density associated to the ordinary matter is absent.
We constructed
a generalized noncommutative analogue to include key ideas of duality and
branch changing as well as gravity-driven acceleration and
kinetic inflation.
%in the standard commutative BD cosmology.
In this manner, we have seen that the power-law scale
factor of the Universe (associated to the commutative case)
is generalized to be multiplied with a time-dependent exponential warp
factor, which is a function of the noncommutative parameter and the momentum associated to the
BD scalar field [see relation~(\ref{NC-a-phi})]. Moreover, in this case, in contrast to the commutative
case, we have observed that the BD scalar field is not in the form of
a simple power function of time, but instead, it is obtained from a more
complicated differential equation, which is found to be an incomplete
gamma function [see differential equation~(\ref{NC-diff-phi})].

In the commutative case, because of the appearance of the integration
constant associated to the momentum conjugate of the BD scalar field in the solutions, our model
can be considered as an extended model of the de Sitter--like space
 and O'Hanlon-Tupper solutions. In the latter, the mentioned integration
constant has an interesting property. Namely, under changing its sign,
a symmetry relates a category of the solutions to its corresponding
counterpart, such that the number of models to be discussed
can be reduced by half. More precisely,
this integration constant together with others presented in
the model can play the role of the duality transformations
introduced in the context of the BD theory~\cite{L95-L96}.

After a short discussion of the consequences of our model within the
standard commutative case, we have focused on noncommutative solutions
(and their interesting interpretations in cosmology), which
give very different results with respect to their corresponding in the commutative case.

In case Ia, we have assumed very small negative values of the noncommutative
parameter $\theta$, positive values of $c_2$, and the lower
sign. We have shown that, unlike the time behavior of the BD
scalar field, the time behaviors of the scale factor,
its speed, and acceleration are very
different in the noncommutative case with respect to the commutative case.
Let us be more concrete. When the BD coupling parameter is
restricted\footnote{We should remind that the behaviors
of the quantities, which have been reported for the noncommutative
case in the range $-3/3<\omega<-4/3$, under some different initial conditions can
also be retrieved for $-4/3<\omega<0$.} to
$-3/2<\omega<0$, the scale factor of the commutative case
always accelerates, while, for the noncommutative case, it
accelerates only for the very early times, and after a
very short time, it turns to give a decelerated universe.
This interesting effect of the noncommutative parameter on the behavior of the scale factor
%may imply a first sign
constitutes a feature of
an appropriate alternative model proposed for an
inflationary model, which can overcome the graceful exit problem.

Furthermore, the mentioned behavior of the scale factor can also be
altered with different allowed values of the parameters present in the model.
More precisely, when different values are taken for $\theta$, $\omega$, and $c_2$,
the time interval, speed, and acceleration of the scale factor
associated to the acceleration phase of the very
early era of the universe also change. We have numerically shown that
the $e$-folding number relates to the amount of $c_2$ and/or $\theta$. Our numerical analysis
show that, for case Ia, the noncommutative minisuperspace model, in which the noncommutative
parameter is a constant, can constitute as a viable
phenomenological model herein, at least for an inflationary epoch, when
$\mid\theta\mid$ takes very small values.

Moreover, in case Ia for late times, contrary to the
commutative model in which the scale factor always accelerates,
we get a zero acceleration epoch for the Universe.
This behavior of the scale factor that is occurring very far from the singularity is guaranteed
by the existence of a constant noncommutative parameter, and it is usually interpreted as
coarse-grained explanation of the quantum gravity footprint.

The horizon problem is the main
shortcoming with the standard cosmology, so, we turned to investigate it
in our model. We have shown that by extending the FLRW
vacuum universe, in the standard BD theory,
by introducing a deformation among the minisuperspace
variables, we can overcome this problem.

By means of numerical diagrams, we have shown that the nominal
as well as sufficient requirements associated to the
inflation can be fully satisfied in our model.
In a kinetic inflation model in the context of the BD
theory with a variable $\omega$ for the commutative case~\cite{Lev95-2},
it was claimed that all the accelerations in the D branch suffer from the well-known graceful exit problem.
 Namely, such a problem has a
 direct relation with the time behavior of the quantity $a\phi^{1/2}$.
 More precisely, if $a\phi^{1/2}$ decreases forever, then a branch change is required.
 Indeed, in the commutative case, the mentioned quantity always
 decreases which is in direct contradiction with the observational
 indications concerning the strength of the gravity as well as the
 expansion of today's Universe. This problem is
 properly solved by the effects of the noncommutative parameter, such that at the very
 early times, exactly the same as the commutative case, $a\phi^{1/2}$
 decreases with the cosmic time, while after
 the phase changing, it starts to increase with time, which is
 in agreement with observation.

 Let us look at the above problem from another perspective.
 If we would like to know whether or not the model also provides inflation in the conformal
 Einstein frame, we must consider not only the behavior of the
 scale factor but also check the behavior of the quantity $a_{_{\rm RI}}\equiv[a(t)l_0]/l_{\rm Pl(t)}$,
 in which $a(t)l_0$~(where $l_0$ is a comoving constant length)
 is any physical length, and
 $l_{\rm Pl}(t)\equiv\sqrt{\frac{\hbar G_{\rm eff}}{c^3}}=\sqrt{\frac{\hbar}{c^3\phi(t)}}$
 is the Planck length, which is not a constant in BD theory.
 If the mentioned conditions were satisfied, then inflation is called real inflation~\cite{Faraoni.book}.
As mentioned, in some models, the Planck length decreases faster than
the scale factor, and, thus, the ratio $a_{_{\rm RI}}$ always decreases, which
is not consistent with the observational data. However, in our model, as we have used the
Planck units, the ratio $a_{_{\rm RI}}$ reduces to the
quantity $a\phi^{1/2}$, and, thus, our model provides real inflation.
However, in the case of the commutative model,
for any $\omega$, particularly for the case $\omega=-1$,
which has been known as pre-big-bang cosmology~\cite{LWC00},
the requirements of the real inflation are not fully satisfied~\cite{C98,C99}.

We should be aware of some shortcomings regarding our noncommutative setting herein.
\begin{itemize}
  \item
In our model, as in other investigations
in the context of the BD theory, to retrieve
the acceleration for the early as well as late time epochs,
the BD coupling parameter takes very small values,
which is in contradiction with Solar System experiments.
Different approaches have been presented to solve the shortcomings
with the cosmological models based on the BD theory, especially the
mentioned problem with $\omega$, see, e.g., Refs.~\cite{BM90,LSB89-PSW08-WW12}.

\item
In this paper, we have confined our discussion to the
noncommutative version of the standard BD theory in the
absence of the ordinary matter, but we
can extend this procedure by adding a matter sector, scalar potential, and/or
assuming a variable BD coupling parameter instead of the constant one.
In addition, we can consider other deformed Poisson brackets instead of the one presented here.\\

\item
Another important point is that we have not tested the
predictions of our model (for the very early Universe) by means of density
and gravitational fluctuations around the FLRW background.
In many investigations, by employing different approaches, the
perturbation of the FLRW background in the BD and generalized scalar-tensor
theories have been studied, see, e.g., \cite{Faraoni.book, HN96-BFG96-M09-CDG13}.
Employing perturbation theory, similar to transformations required for finding
the predictions of the model in the conformal Einstein frame, is crucial, but
performing them in the presence of the noncommutative
parameter is very complicated and in some situations may be impossible.
%\bl{dar morede inke chera NC parameter roye raftare phi
%bel akhas G tasir nadare na baes mishe dar hale hazer ma ba in moshkel movajeh beshim ke
%$G\rightarrow\infty$, harchan kare ma mazayaye khodesh ro dare.be papre haye D. H. Coule ref bedam}
\end{itemize}

\section{ACKNOWLEDGMENTS}
S. M. M. Rasouli is grateful for the support of
Grant No. SFRH/BPD/82479/2011 from the Portuguese
Agency Funda\c c\~{a}o para a Ci\^encia e Tecnologia.
This research work was supported by Grants No.
CERN/FP/123618/2011 and No. PEst-OE/MAT/UI0212/2014.\\


\begin{thebibliography}{99}

\bibitem{BD61}C. Brans and R.H. Dicke, Phys. Rev. \textbf{124}, 925 (1961).
\bibitem{Faraoni.book}V. Faraoni, \textit{Cosmology in Scalar Tensor Gravity} (Kluwer Academic, Dordrecht, 2004).
\bibitem{BP01}N. Banerjee and D. Pavon, Phys. Rev. D \textbf{63}, 043504 (2001).
\bibitem{MC07} A.E. Montenegro, Jr. and S. Carneiro, Classical Quantum Gravity \textbf{24}, 313 (2007).
\bibitem{RFM14}S.M. M. Rasouli, M. Farhoudi and P. V. Moniz, Classical Quantum Gravity \textbf{31}, 115002 (2014).
\bibitem{RFS11-R14}S.M. M. Rasouli, M. Farhoudi and H. R. Sepangi,
Classical Quantum Gravity \textbf{28}, 155004 (2011); S. M. M. Rasouli, Prog. Math. Rel., Gravit. Cosmol. \textbf{60}, 371 (2014).
\bibitem{LS89-SA90-S98-CL11-CFPS12}D. La and P. J. Steinhardt, Phys. Rev. Lett. \textbf{62}, 376 (1989);
P. J. Steinhardt and F. S. Accetta, Phys. Rev. Lett. \textbf{64}, 2740 (1990);
S. Capozziello and M. De Laurentis, Phys. Rep. \textbf{509}, 167 (2011);
T. Clifton, P. G. Ferreira, A. Padilla and C. Skordis, Phys. Rep. \textbf{513}, 1 (2012).
\bibitem{BM90}J. D. Barrow and K. Maeda, Nucl. Phys. \textbf{B341}, 249 (1990).
%\bibitem{S98}M. Susperregi, \textit{Phys. Lett. B} \textbf{440}, 257 (1998).
\bibitem{Lev95}J. J. Levin, Phys. Rev. D \textbf{51}, 462 (1995).
\bibitem{Lev95-2}J. J. Levin, Phys. Rev. D \textbf{51}, 1536 (1995).
\bibitem{BV94}R. Brustein and G. Veneziano, Phys. Lett. B \textbf{329}, 429 (1994).

\bibitem{CDS98-SW99-DN01}A. Connes, M. R. Douglas and A. Schwarz, J. High Energy Phys. \textbf{09}, (1998) 003;
M. R. Douglas and N. A. Nekrasov, Rev. Mod. Phys. \textbf{73}, 977 (2001);
%\bibitem{Pol98}J. Polchinski, \textit{String Theory}, (Cambridge University Press, Cambridge, 1998).
N. Seiberg and E. Witten, J. High Energy Phys. \textbf{09}, (1999) 032.

\bibitem{GORS03-GORS03-2-ADMW06-EGOR08}H. Garcia-Compean, O. Obregon, C. Ramirez and M. Sabido, Phys. Rev. D \textbf{68}, 044015 (2003);
H.Garcia-Compean,O. Obregon,C.Ramirez and M.Sabido, Phys. Rev. D \textbf{68}, 045010 (2003);
P. Aschieri, M. Dimitrijevic, F. Meyer and J. Wess, Classical Quantum Gravity \textbf{23}, 1883 (2006);
S. Estrada-Jimenez, H. Garcia-Compean, O. Obregon and C. Ramirez, Phys. Rev. D \textbf{78}, 124008 (2008).

\bibitem{BP04-PM05-AAOSS07-GSS07}G.D. Barbosa and N. Pinto-Neto, Phys. Rev. D \textbf{70}, 103512 (2004);
L. O. Pimentel and C. Mora, Gen. Relativ. Gravit. \textbf{37}, 817 (2005);
 M. Aguero, J. A. Aguilar, S. C. Ortiz, M. Sabido and J. Socorro, Int. J. Theor. Phys. \textbf{46}, 2928 (2007);
  W. Guzman, M. Sabido and J. Socorro, Phys. Rev. D \textbf{76}, 087302 (2007).


\bibitem{RFK11}S. M. M. Rasouli, M. Farhoudi and N. Khosravi, Gen. Rel. Grav. \textbf{43}, 2895 (2011).
\bibitem{RZMM14} S. M. M. Rasouli, A. H. Ziaie, J. Marto, and P. V. Moniz  Phys. Rev. D \textbf{89} 044028 (2014).
\bibitem{GSS11}W. Guzm\'{a}n, M. Sabido and J. Socorro, Phys. Lett. B \textbf{697}, 271 (2011).
\bibitem{Far09}V. Faraoni, Classical Quantum Gravity \textbf{26}, 145014 (2009).

\bibitem{Jordan55-FGN99}P. Jordan, \emph{Projective Relativity} (Friedrich Vieweg und Sohn, Braunschweig, 1955);
V. Faraoni, E. Gunzig and P. Nardone, Fundam. Cosm. Phys. \textbf{20}, 121 (1999).

\bibitem{C98}D. H. Coule, Classical Quantum Gravity \textbf{15}, 2803 (1998).

\bibitem{BKM04-DDB07-BS07-B09}J.D. Barrow, D. Kimberly and
J. Magueijo, Classical Quantum Gravity \textbf{21}, 4289 (2004);
M.P. Dabrowski, T. Denkiewicz and D. Blaschke, Ann. Phys. (Amsterdam) \textbf{16}, 237 (2007);
K.A. Bronnikov and A.A. Starobinsky, JETP Lett. \textbf{85}, 1 (2007);
P. Bonifacio, Ph.D. thesis, University of Aberdeen, 2009, arXive: gr-qc/0906.0463.



\bibitem{COR02}H. Garcia-Compean, O. Obregon and C. Ramirez, Phys. Rev. Lett. \textbf{88}, 161301 (2002).
\bibitem{L97}J. E. Lidsey, Phys. Rev. D \textbf{55}, 3303 (1997).

\bibitem{o'hanlon-tupper-72-KE95-MW95}J. O'Hanlon and B.O.J. Tupper, Nuovo Cimento Soc. Ital. Fis. \textbf{7B}, 305 (1972);
S.J. Kolitch, and D.M. Eardley, Ann. Phys. (N.Y.) \textbf{241}, 128 (1995);
J.P. Mimoso and D. Wands, Phys. Rev. D \textbf{51}, 477 (1995).



\bibitem{V91-GV92-TV92-GPR94}G. Veneziano, Phys. Lett. B \textbf{265}, 287 (1991);
M. Gasperini and G. Veneziano, Phys. Lett. B \textbf{277}, 256 (1992);
A. A. Tseytlin and C. Vafa, Nucl. Phys. B \textbf{372}, 443 (1992);
A. Giveon, M. Porrati, and E. Rabinovici, Phys. Rep. \textbf{244}, 77 (1994).

\bibitem{L95-L96}J. E. Lidsey, Phys. Rev. D \textbf{52}, R5407 (1995); Classical Quantum Gravity \textbf{13}, 2449 (1996).

\bibitem{F11} V. Faraoni, Phys. Rev. D \textbf{84}, 024003 (2011).
\bibitem{HTW94} Y. Hu, M. S. Turner and E. J. Weinberg, Phys. Rev. D \textbf{49}, 3830 (1994).
\bibitem{LF94}J. J. Levin and K. Freese, Nucl. Phys. \textbf{B421}, 635 (1994).
\bibitem{GOR02}H. Garcia-Compean, O. Obregon and C. Ramirez, Phys. Rev. Lett. \textbf{88}, 161301 (2002).
\bibitem{LWC00} J. E. Lidsey, D. Wands and E. J. Copeland, Phys. Rep. \textbf{337}, 343 (2000).
\bibitem{C99}D. H. Coule, Phys. Lett. B \textbf{450}, 48 (1999).

\bibitem{LSB89-PSW08-WW12}D. La, P. J. Steinhardt and E. Bertschinger, Phys. Lett. B \textbf{231}, 231 (1989);
%\bibitem{BM90}J. D. Barrow and K. Maeda, \textit{Nucl. Phys. B} \textbf{341}, 249 (1990).
R. Punzi, F. P. Schuller and M. N. R. Wohlfarth, Phys. Lett. B \textbf{670}, 161 (2008);
Yu-Huei Wu and Chih-Hung Wang, Phys. Rev. D \textbf{86}, 123519 (2012).


\bibitem{HN96-BFG96-M09-CDG13} Jai-chan Hwang and Hyerim Noh, Phys. Rev. D \textbf{54}, 1460 (1996);
J. P. Baptista, J. C. Fabris and S. V. B. Gonçalves, Astrophys. Space Sci. \textbf{246}, 315 (1996);
T. Matsuda, Classical Quantum Gravity \textbf{26}, 145016 (2009);
J. A. R. Cembranos, A. de la Cruz Dombriz, and L. O. Garc\'{i}a, Phys. Rev. D \textbf{88}, 123507 (2013).


%%%%%%%%%%%%%%%%%%%%%%%%%%%%%%%%%%%%%%%%%%%%%%%%%%%%%%%%%%%%%%%%%%%%%%%%%%%%%%%%%%%%%%%%%%%%%%%%%%%%%%%%%%%%%%%%%%%%%%%%%%%%%%%%%%%%%%%%
%%%%%%%%%%%%%%%%%%%%%%%%%%%%%%%%%%%%%%%%%%%%%%%%%%%%%%%%%%%%%%%%%%%%%%%%%%%%%%%%%%%%%%%%%%%%%%%%%%%%%%%%%%%%%%%%%%%%%%%%%%%%%%%%%%%%%%%%
%%%%%%%%%%%%%%%%%%%%%%%%%%%%%%%%%%%%%%%%%%%%%%%%%%%%%%%%%%%%%%%%%%%%%%%%%%%%%%%%%%%%%%%%%%%%%%%%%%%%%%%%%%%%%%%%%%%%%%%%%%%%%%%%%%%%%%%%
%%%%%%%%%%%%%%%%%%%%%%%%%%%%%%%%%%%%%%%%%%%%%%%%%%%%%%%%%%%%%%%%%%%%%%%%%%%%%%%%%%%%%%%%%%%%%%%%%%%%%%%%%%%%%%%%%%%%%%%%%%%%%%%%%%%%%%%%
%%%%%%%%%%%%%%%%%%%%%%%%%%%%%%%%%%%%%%%%%%%%%%%%%%%%%%%%%%%%%%%%%%%%%%%%%%%%%%%%%%%%%%%%%%%%%%%%%%%%%%%%%%%%%%%%%%%%%%%%%%%%%%%%%%%%%%%%

\end{thebibliography}
\end{document}